\documentclass[preprintnumbers,amsmath,pacs,amssymb]{revtex4}
\usepackage{amsmath}
\usepackage{yfonts}
\usepackage{marvosym}
\usepackage{extarrows}
\usepackage[ruled]{algorithm2e}
\usepackage{graphicx}
\usepackage{dcolumn}
\usepackage{bm}
\usepackage[all]{xy}
\usepackage{indentfirst}
\usepackage{amsmath}
\usepackage{multirow}
\usepackage{mathrsfs}
\usepackage{bbm}
\usepackage{euscript}
\usepackage{amssymb}
\begin{document}
\title{Nonlocality of All Quantum Networks}
 \author{Ming-Xing Luo}
\affiliation{\small Information Security and National Computing Grid Laboratory,
\\
\small Southwest Jiaotong University, Chengdu 610031, China}

\begin{abstract}
The multipartite correlations derived from local measurements on some composite quantum systems are inconsistent with those reproduced classically. This inconsistency is known as quantum nonlocality and shows a milestone in the foundations of quantum theory. Still, it is NP hard to decide a nonlocal quantum state. We investigate an extended question: how to characterize the nonlocal properties of quantum states that are distributed and measured in networks. We first prove the generic tripartite nonlocality of chain-shaped quantum networks using semiquantum nonlocal games. We then introduce a new approach to prove the generic activated nonlocality as a result of entanglement swapping for all bipartite entangled states. The result is further applied to show the multipartite nonlocality and activated nonlocality for all nontrivial quantum networks consisting of any entangled states. Our results provide the nonlocality witnesses and quantum superiorities of all connected quantum networks or nontrivial hybrid networks in contrast to classical networks.
\end{abstract}

\maketitle

\section{Introduction}

The joint probability distribution of the outcomes of local measurements performed on spatially separated quantum systems sometimes exhibit correlations that cannot be explained classically in terms of the information shared beforehand. Such nonlocal correlations are revealed by the violation of special inequalities [1-3]. Formally, Bell's theorem [3] states that the predictions of quantum mechanics are inconsistent with classical causal relations those originate from a classical common local hidden variable (LHV). The profound property holds for general quantum entangled systems [4-12] whose joint state cannot be written in a mixture of states in product forms. These typical correlations are found of great interest in information processing \cite{GH,Masa,PAMB}, communication \cite{Ek,SG,MY,AGM,ABG}, quantum theory and potential applications \cite{BPA,BCMD,BCPS}.

Quantum entanglement is a valuable resource for various tasks including quantum key distribution, randomness extraction, and quantum communication \cite{HHH0,CSS}. Interestingly, different from classical states entangled states can be swapped \cite{ZZHE,SPG}, where a local measurement can create an entanglement for two parties who have no prior shared entanglement as shown in Figure 1. The remarkable feature is the foundation of distributing quantum entanglement in long distance \cite{TRO,TBZ,AB,YC} and constructing large-scale quantum networks \cite{Kimb,Ritt,BDCZ,DLCZ,ATL,ZPD}. Nonetheless, it still remains an open problem to characterize these quantum behaviors, even if for the simplest quantum network consisting of two bipartite entangled states.

\begin{figure}
\begin{center}
\includegraphics[width=.4\textwidth]{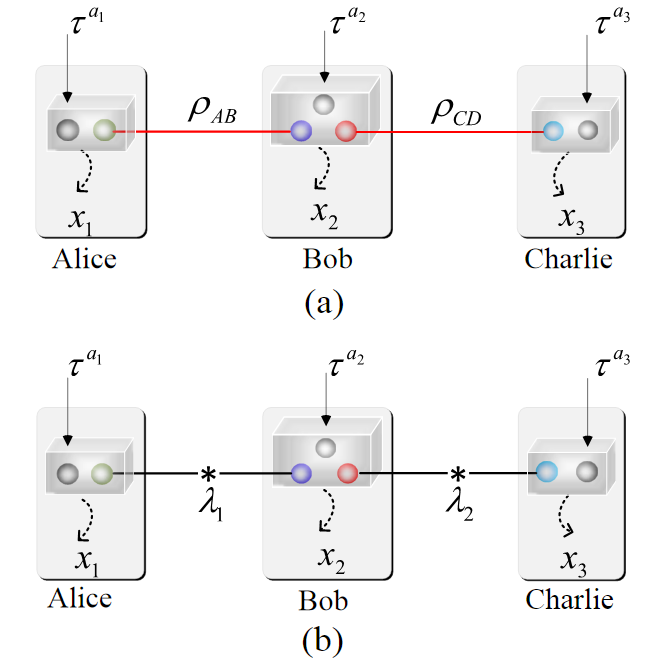}
\end{center}
\caption{(Color online) Schematic Bell testing of quantum entanglement swapping using a generalized Bell inequality \cite{SQ}. (a) Tripartite Bell testing of two bipartite entangled states $\rho_{AB}\otimes \rho_{CD}$. $\tau^{a_1}, \tau^{a_2}, \tau^{a_3}$ are input states of Alice, Bob, and Charlie, respectively. $x_1, x_2, x_3$ are measurement outcomes of three parties. (b) Classical hidden state model for testing the locality of two shared sources $\lambda_1, \lambda_2$ or separable states $\rho_{AB}=\sum_{i}p_i\varrho^{(i)}_{A}\otimes \varrho^{(i)}_{B}$ and $\rho_{CD}=\sum_{i}q_i\varrho^{(i)}_{C}\otimes \varrho^{(i)}_{D}$, where $\{p_i\}$ and $\{q_i\}$ are probability distributions and $\varrho^{(i)}_{A(B,C,D)}$ are density operators of system $A$ ($B, C, D$). The probabilities of $\tau^{a_i}$s are uniform distributions.}
\end{figure}

In comparison to single entanglement, there are independent sources in a general quantum network for distributing hidden states to space-like separated parties in terms of the locally causal model \cite{Bell,Cire,PR,ABH}. The standard Bell testing \cite{Bell,CHSH} is useless for verifying these entangled states in a distributive model. Recently, nonlinear Bell-type inequalities are proposed for verifying the non-trilocality \cite{CAS} or non-bilocality \cite{BGP,GMT} of correlations originating from the standard entanglement swapping. It is then extended for star-shaped networks \cite{ACS} or small-sized networks \cite{LS,CH}. Another procedure is iteratively expanding a given network into the desired network \cite{RBB}. Different from these methods, a polynomial-time algorithm is proposed for constructing explicit nonlinear Bell-type inequalities for general networks \cite{Luo}. Their inequalities are useful for proving the generic non-multilocality of quantum networks consisting of all bipartite entangled pure states \cite{EPR1} and generalized Greenberger-Horne-Zeilinger (GHZ) states \cite{GHZ}. Certain quantum experiments are performed for verifying the nonlocalities \cite{CL,BG,TWE,SBB,CASB}. Unfortunately, there is no result to feature all quantum networks.

Our goal in this work is to prove the nonlocality of all quantum networks consisting of any entangled states, which is a weak problem of verifying  all quantum networks. We specially investigate the existence of generalized Bell testing for verifying multipartite correlations generated by local measurements in each nontrivial quantum network in terms of the generalized locally causal model \cite{PR,Cire}. Our approach depends primarily on the recent semiquantum nonlocal game \cite{SQ} that permits to verify single entanglement. We firstly prove that tripartite correlations of the quantum network shown in Figure 1 violate generalized Bell inequalities for all bipartite entangled states. The generic non-multilocality is different from the nonlocality of single entanglement using CHSH inequality \cite{CHSH,GP,PR,LF} or Hardy inequality \cite{Hardy,YCZ}. Furthermore, the nonlocality of two independent parties without initially sharing entanglement can be activated by a local measurement of the other party. It is of another remarkable feature of entangled states \cite{SPG} and provides a way for detecting a single entanglement. These results are then applied to the multipartite nonlocality of all connected quantum networks consisting of any entangled states, and the activated nonlocality for any subnetwork consisting of independent parties. Finally, we show the multipartite nonlocality of nontrivial hybrid networks consisting of entangled states and classical resources. Remarkably, the result holds for any non-classical network and provides a quantum superiority of hybrid Internet over all classical networks.

\section{Results}

{\it Multilocality structure of a network}. Consider a network consisting of classical systems that are shared by $n$ parties, $\texttt{P}_1$, $\texttt{P}_2$, $\cdots$, $\texttt{P}_n$. In experiment, each party  $\texttt{P}_i$ perform $m$ local measurements on their respective subsystems with $k$ possible outcomes, $x_i=1, \cdots, k$, where $k>2$ for general systems \cite{Luo1}. Denote $a_i=1, \cdots, m$ as the measurement chosen by $\texttt{P}_i$. The joint conditional probability distribution $P({\bf x}|{\bf a})$ of all measurement outcomes with ${\bf x}=x_1\cdots x_n $ and ${\bf a}=a_1\cdots a_n$ is local whenever it can be explained as the result of classically correlated datas, represented by hidden variables $\lambda_1, \cdots, \lambda_s$, that is,
\begin{eqnarray}
P_{local}({\bf x}|{\bf a})=\int_{\Omega} d\mu(\lambda_1)\cdots d\mu(\lambda_s) \prod_{j=1}^np(x_j|a_j,\Lambda_j),
\label{eqn1}
\end{eqnarray}
where $\Lambda_j$ contains some variables $\lambda_i$s, $\mu_i(\lambda_i)$ denotes the measure of $\lambda_i$ with the normalization condition $\int_{\Omega_i} d\mu_i(\lambda_i)=1$ and $(\Omega_i, \Sigma_i, \mu_i)$ is the measure space of $\lambda_i$, $i=1, 2, \cdots, s$. Every local distribution (\ref{eqn1}) satisfies special constraint known as Bell-type  inequality \cite{Luo}.

Local measurements on some composite quantum states $\rho_1, \cdots, \rho_k$ in a distributive model or network lead to the violation of proper Bell inequality that is then a signature of the non-multilocality \cite{BGP,CH,GMT,RBB,Luo}. Specifically, there exist joint conditional probability distributions resulting from local quantum measurement outcomes of all observers as
\begin{eqnarray}
P_Q({\bf x}|{\bf a})=
\textrm{Tr}[(\otimes_{i=1}^n{M}^{x_i}_{a_i})(\otimes_{j=1}^k\rho_j)],ªí
\label{eqn2}
\end{eqnarray}
which cannot be reproduced by any local model (\ref{eqn1}), where $\{{M}^{x_i}_{a_i}, \forall x_i\}$ are positive operators describing  locally implementable quantum measurements by the observer $\mathtt{P}_i$ and satisfy $\sum_{x_i}M^{x_i}_{a_i}=\mathbbm{1}$ for each $a_i$, and $\mathbbm{1}$ is the identity operator. The joint quantum system $\rho_1\otimes \cdots \otimes\rho_k$ is multipartite nonlocal and reduces to Bell's nonlocal for $k=1$ \cite{Bell}. A stronger nonlocality is possible in hidden state scenario with quantum inputs \cite{SQ}. $n$ random sources $\{\tau^{a_1}, \forall a_1\}, \cdots, \{\tau^{a_n}, \forall a_n\}$ are assumed for all observers shown in Figure 1. It follows a new joint conditional probability distribution as
\begin{eqnarray}
{P}_Q({\bf x}|{\bf a})\!\!\!\!\!\!&&={P}_Q({\bf x}|\tau^{a_1}, \cdots, \tau^{a_n})
\nonumber\\
\!\!\!\!\!\!&&=\textrm{Tr}[(\otimes_{i=1}^n{M}^{x_i}_{a_i})((\otimes_{j=1}^k\rho_j)\otimes (\otimes_{s=1}^n\tau^{a_s}))].ªí
\label{eqn2}
\end{eqnarray}
Similar to verifying single entanglement \cite{Bell,CHSH}, how to decide the nonlocality of a quantum network is also a fundamental question. Nevertheless, identifying the nonlocality of a general quantum network remains an extremely difficult problem [40-47].

Here, we introduce a new framework for exploring the nonlocality of all quantum networks. Given a quantum network consisting of a joint quantum system $\rho=\rho_1\otimes \cdots\otimes \rho_k$ shared by $n$ observers, the main idea is to create quantum subnetworks for special observers (Alice and Charlie shown in Figure 1). If some nonlocal correlations can be observed in these subnetworks, they are provided by the quantum state $\rho$, which is then multipartite nonlocal. Within the new scenario, we investigate the activation phenomena for these subnetworks consisting of all independent observers without prior sharing entanglement. It is of the multipartite nonlocality in a network scenario.

\textbf{Definition 1}. A quantum network ${\cal N}_q$ is multipartite nonlocal if a set of observables existing for all observers such that multipartite quantum correlations from local measurements are inconsistent with these from the generalized local realism.

\textbf{Definition 2}. A quantum network ${\cal N}_q$ consisting of $n$ observers is $k$-partite activated nonlocal if for any $s$ observers $\texttt{p}_{i_1}, \cdots, \texttt{p}_{i_s}$ with $2\leq s\leq k$, there is a set of observables for all observers such that a local measurement of $n-s$ observers $\texttt{p}_{j}$s with $j\in \{1, \cdots, n\}\backslash\{i_1, \cdots, i_s\}$ creates a $s$-partite nonlocal subnetwork.

\textbf{Nonlocality of all $\Lambda$-shaped quantum networks}. Consider a $\Lambda$-shaped quantum network ${\cal N}_q$ consisting of three observers Alice, Bob and Charlie shown in Figure 1. The nontrivial feature of ${\cal N}_q$ is entanglement swapping \cite{ZZHE,BGP}. Different from the standard nonlocality \cite{GP,PR,Gisin1,Pop} detected by linear Bell inequalities \cite{Bell,CHSH}, the tripartite nonlocality of ${\cal N}_q$ can be verified using nonlinear Bell-type inequalities for all bipartite entangled pure states and special entangled mixed states \cite{GMT,Luo}. Our goal here is to prove the tripartite nonlocality and bipartite activated nonlocality of ${\cal N}_q$ for all bipartite entangled states. Specifically, assume that ${\cal N}_q$ consists of two bipartite entangled states $\rho_{AB}$ and $\rho_{CD}$, where Alice has particle $A$, Bob has particles $B$ and $C$ while Charlie has particle $D$. The joint state of ${\cal N}_q$ reading $\rho=\rho_{AB}\otimes \rho_{CD}$ is activated nonlocal for all entangled pure states \cite{GMT,Luo}. Our first result is to generalize the result to all bipartite entangled states. Formally, we prove the following result

\textbf{Theorem 1}. Assume that a $\Lambda$-shaped quantum network ${\cal N}_q$ consists of any bipartite entangled states shared by three observers. Then the following results hold: (i) ${\cal N}_q$ is tripartite nonlocal; (ii) ${\cal N}_q$ is bipartite activated nonlocal.

Different from previous Bell inequalities for verifying special $\Lambda$-shaped quantum networks \cite{ZZHE,BGP,GMT,Luo}, Theorem A shows that a generalized Bell-type inequality exists for a given $\Lambda$-shaped quantum network consisting of any bipartite entangled states. Generally, this Bell-type inequality is state-dependent. Furthermore, one measurement of Bob can create an entanglement between Alice and Charlie who have no prior shared entanglement. The interesting feature of bipartite entangled states is going beyond classical resources \cite{SPG} and key to build large-scale quantum networks \cite{Kimb,Ritt,ZPD}. The proof of the tripartite nonlocality stated in Theorem A is a straight forward application of the semiquantum nonlocal game for each entanglement \cite{SQ}. For the bipartite activated nonlocality, our proof will be completed for a reduced quantum network consisting of qubit-based entangled states. The main idea is that Alice and Bob are allowed to firstly perform local projections on high-dimensional systems and local distilling of entangled mixed states \cite{HHH,SI}. These assumptions are reasonable because local operations and classical communication (LOCC) or local operations and shared randomness (LOSR) cannot create entanglement between two observers initially sharing no entanglement \cite{HHH,SQ}. Additionally, the proof of Theorem 1 also provided an interesting by-product that universal Bell inequality exists for detecting a single entanglement by using local projection and entanglement distilling \cite{SQ}.

\begin{figure}
\begin{center}
\includegraphics[width=.5\textwidth]{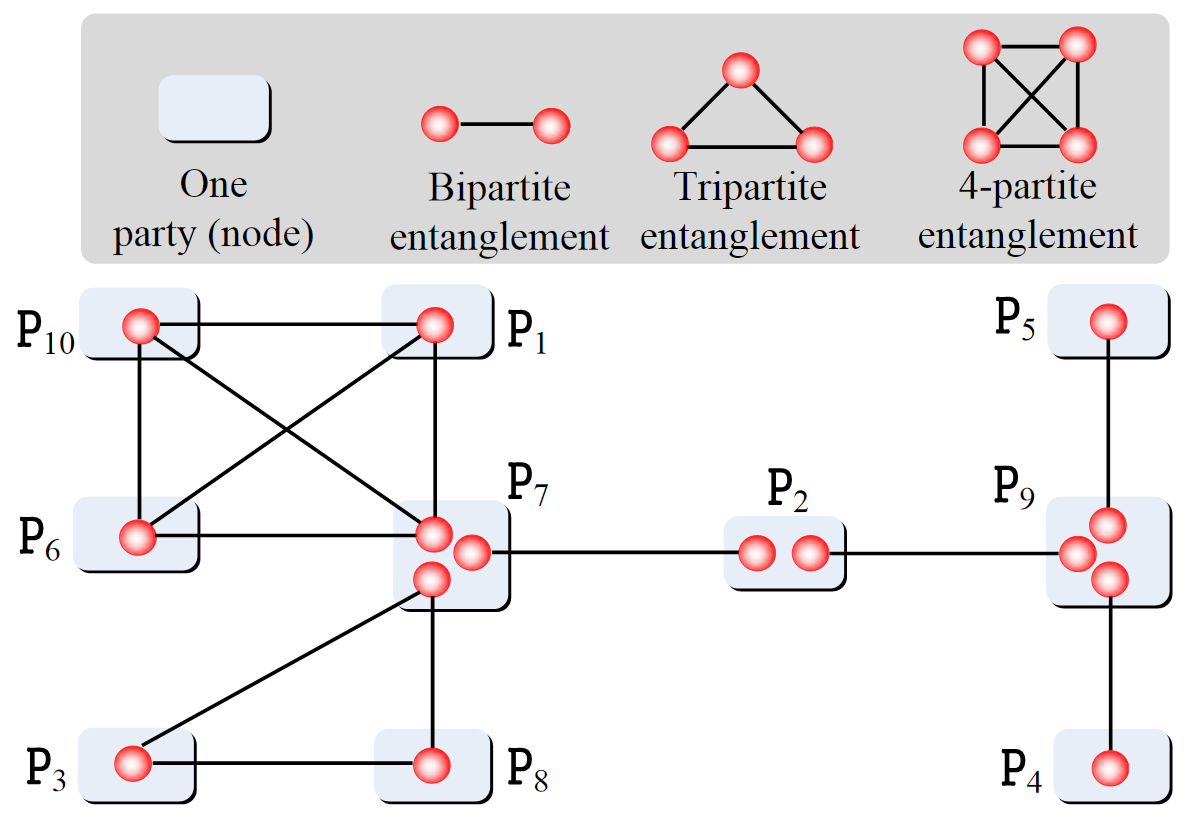}
\end{center}
\caption{(Color online) A schematic connected quantum network. Quantum resources consist of 4 bipartite entangled states, one tripartite entanglement and one 4-partite entanglement that are shared by 10 observers $\texttt{P}_1, \cdots, \texttt{P}_{10}$. $\texttt{P}_i$ and $\texttt{P}_{9}$ share one bipartite entanglement for $i=2, 4, 5$. $\texttt{P}_2$ and $\texttt{P}_{7}$ share one bipartite entanglement. $\texttt{P}_3, \texttt{P}_7$ and $\texttt{P}_{8}$ share one tripartite entanglement. $\texttt{P}_1, \texttt{P}_6, \texttt{P}_7$ and $\texttt{P}_{10}$ share one 4-partite entanglement. Each ball denotes one physical particle.}
\end{figure}

\textbf{Nonlocality of all connected quantum networks}. The $\Lambda$-shaped quantum network can be generalized to chain-shaped quantum networks \cite{RBB,Luo} or star-shaped quantum networks with several observers \cite{ACS,RBB,Luo}. Here, we further consider a general quantum network ${\cal N}_q$ with the connectivity in its schematic graph scenario. Our goal is to prove the multipartite nonlocality and activated locality of these networks. Specifically, assume that ${\cal N}_q$ is connected network if for any two observers $\texttt{P}_k$ and $\texttt{P}_s$ there is a set of observers $\texttt{P}_{i_1}, \cdots, \texttt{P}_{i_t}$ such that any adjacent two observers of $\texttt{P}_k, \texttt{P}_{i_1}, \cdots, \texttt{P}_{i_t}, \texttt{P}_s$ share at least one entanglement. Otherwise, ${\cal N}_q$ is disconnected. The definition can be also defined as the connectivity of its equivalent graph \cite{BLW}, where each $s$-partite entanglement is schematically represented by a complete graph with $s$ distinct vertices, and each node represents an observer shown in Figure 2. Our applications in the following are all connected quantum networks. Similar to Theorem 1, we can prove the multipartite nonlocality and activated nonlocality for generic connected quantum networks. It is formally stated as the following result

\textbf{Theorem 2}. Assume that a connected quantum network ${\cal N}_q$ consists of any multipartite entangled states shared by $n$ observers. Then the following results hold: (i) ${\cal N}_q$ is multipartite nonlocal; (ii) ${\cal N}_q$ is $k$-partite activated nonlocal for any $2\leq k\leq n-1$.

In contrast to the genuine multipartite nonlocality of special quantum networks \cite{SS,CGP} detected by Svetlichny inequalities \cite{Sve} or multipartite nonlocality detected by nonlinear Bell-type inequalities \cite{BGP,GMT,Luo}, Theorem 2 provides the existence of generalized Bell-type inequalities for verifying each connected quantum network. The result is further extended to any subnetwork with the activated nonlocality. The surprising part is similar to the standard entanglement swapping that permits some observers in ${\cal N}_q$ help independent  observers to construct nonlocal quantum correlations going beyond classical scenario \cite{SPG}. Theorem 2 implies that this interesting feature is generic for all connected quantum networks consisting of any entangled states going beyond chain-shaped or star-shaped quantum networks \cite{ACS,RBB,Luo}. It is special important for constructing general quantum networks. Similar to Theorem 1, the first proof of Theorem 2 is an application of the semiquantum nonlocal game \cite{SQ,SI}. To prove the activated nonlocality, an iterative algorithm will be proposed to reduce a general quantum network into a hybrid network consisting of several chain-shaped and star-shaped quantum subnetworks. The projection method and entanglement distilling are then used to complete the proof \cite{HHH,SI}.

\begin{figure}
\begin{center}
\includegraphics[width=.65\textwidth]{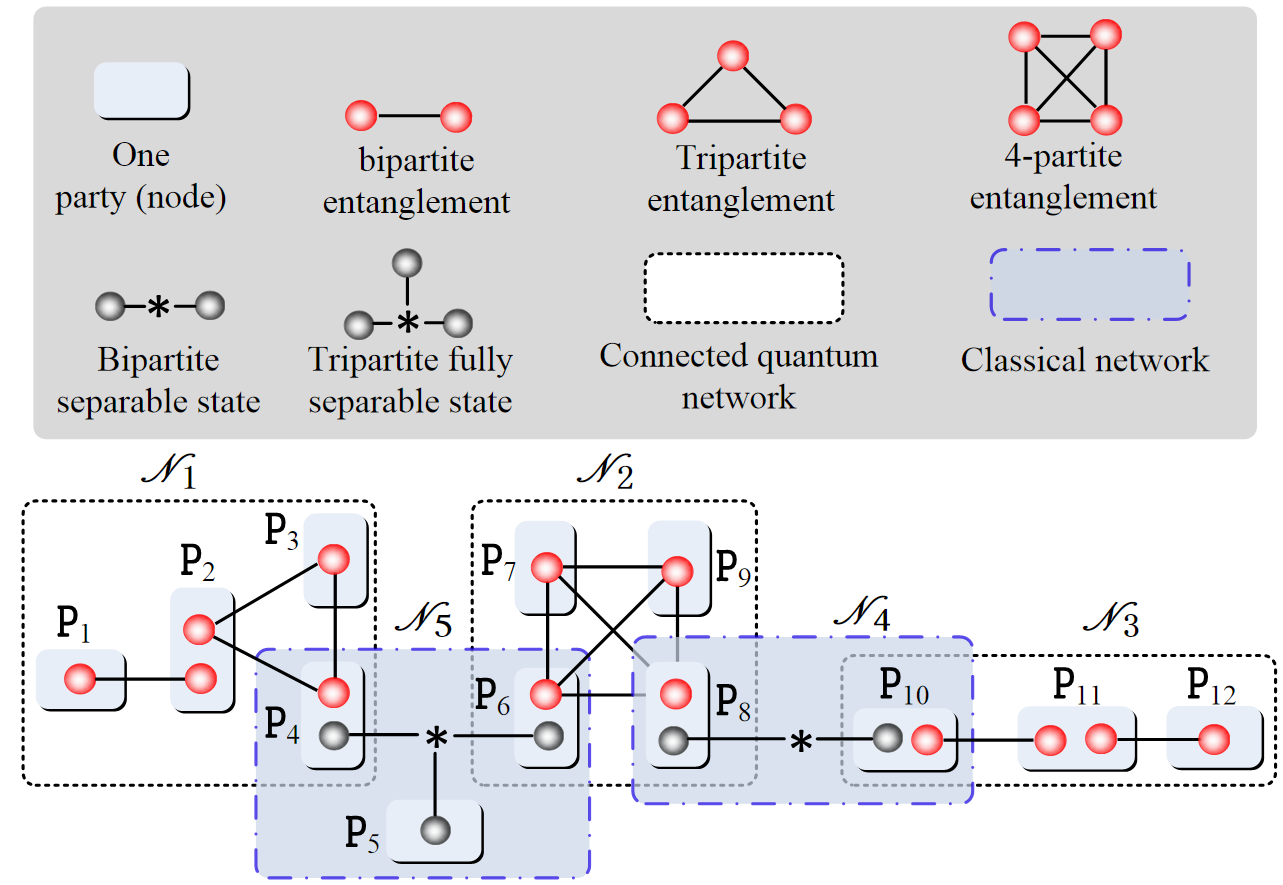}
\end{center}
\caption{(Color online) A schematic hybrid network. There are five subnetworks ${\cal N}_1, \cdots, {\cal N}_5$ consisting of 12 observers $\texttt{P}_1, \cdots, \texttt{P}_{12}$. $\texttt{P}_i$ and $\texttt{P}_{i+1}$ share one bipartite entanglement for $i=1, 10, 11$. $\texttt{P}_2, \texttt{P}_3$ and $\texttt{P}_{4}$ share one tripartite entanglement. $\texttt{P}_6, \cdots, \texttt{P}_9$ share one 4-partite entanglement. $\texttt{P}_4, \texttt{P}_5$ and $\texttt{P}_6$ share one tripartite fully separable state. $\texttt{P}_8$ and $\texttt{P}_{10}$ share one bipartite separable state. ${\cal N}_1, {\cal N}_2$ and ${\cal N}_3$ are connected quantum subnetworks while ${\cal N}_4$ and ${\cal N}_5$ are classical subnetworks. Here, each fully separable state can be equivalently described by a local hidden variable in terms of LHV model.
}
\end{figure}

\textbf{Quantum superiority of all nontrivial hybrid networks}. Theorem 2 provides a strong multipartite nonlocality in terms of the activated nonlocality for all connected quantum networks. An available network can consist of hybrid resources including quantum entangled states and classical resources (or fully separable states) shown in Figure 3. A natural problem states: is it possible to characterize these hybrid networks? So far, no related result has been obtained for this problem. Note that these networks are generally disconnected quantum networks and then cannot provide the activated nonlocality for all subnetworks. Although the multipartite correlations from local measurements of hybrid networks do not satisfy the activated nonlocality given in Definition 2, our goal here is to explore the global quantum nonlocality obtained from these networks. Specifically, is it possible to define a generalized Bell-type inequality that can witness a hybrid network in contrast with classical networks? Different from Theorem 2 we prove weak multipartite nonlocality or global quantum supremacy for hybrid networks as follows:

\textbf{Theorem 3}. Assume that a nontrivial hybrid network ${\cal N}_{cq}$ consist of any multipartite entangled states and fully separable states. Then ${\cal N}_{cq}$ is multipartite nonlocal.

A nontrivial hybrid network in Theorem 3 denotes any hybrid network consisting of at least one multipartite entangled state. Different from the entanglement detection \cite{Bell} or the verification of quantum networks \cite{CAS,BGP,GMT,Luo}, Theorem 3 provides the first generic result of nontrivial hybrid networks. The following equivalent interpretation of Theorem 3 is more interesting. The global nonlocality of ${\cal N}_{cq}$ can be used to acquire quantum superiority over all classical networks in terms of the collaborative semiquantum game \cite{SQ,SI}, where without communicating with each other all players maximize a negative payoff using local measurement outcomes. Especially, from Theorem 3 for all players there is a general payoff $\wp$ depending on multipartite correlations $p(\textbf{x}|\textbf{a})$ satisfying a strictly inequality $\wp_{cq}>\wp_{c}$, where $\wp_{cq}$ and $\wp_{c}$ are maximal payoffs using multipartite correlations of ${\cal N}_{cq}$ and all classical networks, respectively. It suggests an interesting way to design  distributive tasks for any hybrid network \cite{Kimb}. More importantly, it provides a quantum superiority of each nontrivial hybrid network over all classical networks.

\section*{Discussions}

Different from previous multipartite nonlocality using nonlinear Bell-type inequalities \cite{BGP,CH,GMT,RBB,Luo}, Definition 1 provides the weakest version to show a global nonlocality of quantum entangled states in  network scenario \cite{CAS}, where one entanglement can imply the global nonlocality from Theorem 3. Definition 2 can be described as an extension of the genuine multipartite nonlocality of single entanglement \cite{Sve,SS,CGP} for a quantum network consisting of multiple entangled states. In the strong statement, the multipartite correlations $P({\bf x}|{\bf a})$ cannot be decomposed into a convex combination of $s$-partite with $1\leq s\leq n-1$ and $n-s$ partite correlations with respect to any fixed bipartite partition of $n$ observers, i.e., $P({\bf x}|{\bf a})\not=\sum_{\lambda}p(\lambda)P({\bf x}_s|{\bf a}_k,\lambda)P({\bf x}_{n-s}|{\bf a}_{n-s},\lambda)$, where ${\bf x}_{s}=x_{i_1}\cdots x_{i_s}$, ${\bf a}_{s}=a_{i_1}\cdots a_{i_s}$, ${\bf x}_{n-s}$ and ${\bf a}_{n-s}$ denote all the other variables of $x_i$s and $a_i$s except for ${\bf x}_{s}$ and ${\bf a}_{s}$, respectively. Theorems 1 and 2 imply that all connected quantum networks consisting of $n$ observers are genuine multipartite nonlocal, where $2\leq k \leq n-1$. This result can be easily followed from the definition of connected quantum networks.

A weak version of Definition 2 for a quantum network ${\cal N}_q$ is to consider the activated nonlocality of special subnetworks. In this case, the activated nonlocality does not hold for all subnetworks. Formally, ${\cal N}_q$ is $k$-partite weak activated nonlocal if for some independent $k$ observers there exist local measurements of all the other $n-k$ observers such that one of their outcomes creates a nonlocal subnetwork. The special case is also useful for special goals. From new definition, the weak activated nonlocality can be proved for some nontrivial hybrid networks with connected subnetworks, where LOCC or LOSR cannot create entanglement between two parties initially sharing no entanglement \cite{HHH,SQ}. Interestingly, recent results \cite{CHW,BRL} provide a method to construct Bell-type inequalities for single entanglement based on its entanglement witness. Unfortunately, there is no explicit algorithm to construct the entanglement witness for all entangled states. New explorations should be interesting for quantum many-body systems \cite{Deng,LGLS,ACTC}. Besides, it is unknown how to verify general entanglement swapping without entanglement distilling \cite{HHH}. We provide some further discussions by classifying entangled mixed states \cite{SQ}.

We showed that any $\Lambda$-shaped quantum network is tripartite nonlocal and bipartite activated nonlocal, which provided an evidence of generic entanglement swapping for all entangled states. As a nontrivial application, we obtained a generalized result that the multipartite nonlocality holds for all connected quantum network consisting of any multipartite entangled states. The interesting part is the activated nonlocality of these quantum networks, which is a natural extension of the genuine multipartite nonlocality of single entanglement. As another by-product, each nontrivial hybrid network can provide quantum superiority over all classical networks in terms of the semiquantum nonlocal game, in which the joint question-answer probability distributions of a nontrivial hybrid network cannot be described classically with any shared randomness. These results show the usefulness of quantum networks including hybrid Internet going beyond classical networks.

\section*{Acknowledgements}

We thank the helpful discussions of Luming Duan. This work is supported by the National Natural Science Foundation of China (Grants No.61772437, 61702427), Sichuan Youth Science and Technique Foundation (Grant No.2017JQ0048), Fundamental Research Funds for the Central Universities (Grant No. 2682014CX095), and financial support from Chuying fellowship.

\newpage

\begin{center}
{\Large \bf Supplementary Information ``Nonlocality of All Quantum Networks"}
\vskip 0.7cm
{\large Ming-Xing Luo}
\\
{\it \small Information Security and National Computing Grid Laboratory,}
\\
{\it \small Southwest Jiaotong University, Chengdu 610031, China}
\end{center}

\setcounter{figure}{0}
\newcommand{\reffig}[1]{Figure \ref{#1}}
\renewcommand \thefigure{S\arabic{figure}}

\section*{Appendix A: Proof of Theorem 1}

Before we present the detailed proofs of Theorem 1 and the other results, we introduce some necessary definitions inspired by the semiquantum nonlocal game \cite{SQ} to rigorously state the main result. In what follows, all quantum systems are finite dimensions (i.e., their Hilbert spaces, denoted by ${\cal{H}}$, are finite dimensions) and index sets (denoted by ${\cal{S}}, {\cal{T}}, {\cal{X}}$, and ${\cal{Y}}$) contain only a finite number of elements. The convex set of probability distributions defined on an index set ${\cal{X}}$ is denoted by ${\cal{P}}({\cal{X}})$. The set of linear operators acting on a Hilbert space ${\cal{H}}$ is denoted by $\mathbb{L}({\cal{H}})$. The set of density matrices (i.e., positive semidefinite, trace-one operators) is denoted by $\mathbb{S}({\cal{H}})\subset \mathbb{L}({\cal{H}})$.

A random source of states of a quantum system $A$ is represented by an ensemble $\tau=(\{p(s),\tau^s\}; s\in {\cal{S}}\})$, where $p\in {\cal{P}}({\cal{S}})$ and $\tau^s\in\mathbb{S}({\cal{H}}_A)$, for all $s$. Given an outcome set ${\cal{X}}=\{x\}$ and a quantum system $A$ on Hilbert space ${\cal{H}}_A$, an ${\cal{X}}$-probability operator-valued measure (${\cal X}$-POVM, for short) on $A$ is a family $P =(P^x; x\in {\cal{X}})$ of positive semidefinite operators $P^x\in \mathbb{L}({\cal{H}}_A)$, such that $\sum_{x\in {\cal{X}}}P^x=\mathbbm{1}$. We denote by ${\cal{M}}(A; {\cal{X}})$ the convex set of all ${\cal{X}}$-POVMs on $A$. A POVM $P\in {\cal M}(A; {\cal{X}})$ induces, via the relation $p(x)=\textrm{Tr}[P^x\varrho]$, a linear function $P:\varrho\mapsto P\varrho$ from $\mathbb{S}({\cal{H}}_A)$ to ${\cal P}({\cal{X}})$. POVMs in ${\cal M}(A; {\cal{X}})$ are used to represent physical available measurements performed on a quantum system $A$ with outcomes in ${\cal{X}}$.

We firstly verify the following standard $\Lambda$-shaped quantum network consisting of two entangled states. Here, we present detailed statements of Theorem 1 as:

\textbf{Theorem 1}. For any network ${\cal{N}}_q$ consisting of three observers Alice, Bob and Charlie, assume that Alice and Bob share one bipartite entanglement $\rho_{AB}$ on the Hilbert space ${\cal{H}}_{A}\otimes {\cal{H}}_B$, and Bob and Charlie share one bipartite entanglement $\rho_{CD}$ on the Hilbert space ${\cal{H}}_{C}\otimes {\cal{H}}_D$. Then $\rho_{AB}\otimes\rho_{CD}$ is tripartite nonlocal. Moreover, for the subnetwork consisting of Alice and Charlie, there are local observables for three observers such that one local measurement of Bob can create a bipartite entanglement between Alice and Bob, i.e., ${\cal{N}}_q$ is bipartite activated nonlocal.

\textbf{Proof}. Since $\rho_{AB}$ is bipartite entangled,  there is a semiquantum game $\mathbbm{G}_{sq}$ (see Figure S1(a)) \cite{SQ}, constants  $\alpha_{a_1a_2}^{x_1x_2}$, auxiliary states $\tau_{A_0}^{a_1}\in {\cal{H}}_{A_0}$ and ${\mathscr{X}}_1$-POVM $\{P^{x_1}\}\in {\cal{M}}_{A_0A; {\mathscr{X}}_1}$ for Alice, auxiliary states $\tau_{B_0}^{a_2}\in {\cal{H}}_{B_0}$ and ${\mathscr{X}}_2$-POVM $\{Q^{x_2}\}\in {\cal M}_{BB_0; {\mathscr{X}}_2}$ for Bob such that
\begin{align*}
\sum_{a_1,a_2,x_1,x_2}\alpha_{a_1a_2}^{x_1x_2}P(x_1,x_2|a_1,a_2)>c_1
\tag{A1}
\end{align*}
where $P(x_1,x_2|a_1,a_2)$ are joint conditional probability distributions computed as
\begin{align*}
P(x_1,x_2|a_1,a_2)=\textrm{Tr}[(P^{x_1}_{A_0A}\otimes Q^{x_2}_{BB_0})(\tau^{a_1}_{A_0}\otimes \rho_{AB}\otimes \tau^{a_2}_{B_0})],
\tag{A2}
\end{align*}
$c_1$ denotes the maximal achievable classical bound of average gain in terms of the semiquantum game $\mathbbm{G}_{sq}$. Denote ${\mathscr{A}}_i=\{a_i\}$ and ${\mathscr{X}}_i=\{x_i\}$ with $i=1, 2$. Here, we assume that the probability distributions of input random sources $\{\tau_A^{a_1}\}$ and $\{\tau_{B}^{a_2}\}$ are uniform. Otherwise, one can redefine the constants $\alpha_{a_1a_2}^{x_1x_2}$. For all joint conditional probability distributions $P_c(x_1,x_2|a_1,a_2)$s derived from shared classical correlations (hidden variable model \cite{Bell,PR}) or fully separable quantum states (hidden state model \cite{Cire}, see Figure S1(b)), we have
\begin{align*}
\sum_{a_1,a_2,x_1,x_2}\alpha_{a_1a_2}^{x_1x_2}P_c(x_1,x_2|a_1,a_2)\leq c_1,
\tag{A3}
\end{align*}
where $P_c(x_1,x_2|a_1,a_2)$ can be any convex combination of independent distributions in the variables $x_1, x_2$, i.e, $P_c(x_1,x_2|a_1,a_2)=\sum_{j}p_jP_j(x_1|a_1)P_j(x_2|a_2)$, and $\{p_j\}$ is a probability distribution.

Note that the inequalities (A1) and (A3) can be regarded as generalized Bell-type inequalities for testing bipartite entanglement $\rho_{AB}$.

Similarly, for the bipartite entanglement $\rho_{CD}$ there is another semiquantum game $\mathbbm{G}'_{sq}$ (see Figure S1(c)), constants $\beta_{a_3a_4}^{x_3x_4}$, auxiliary states $\tau_{C_0}^{a_3}\in {\cal{H}}_{C_0}$ and ${\mathscr{X}}_3$-POVM $\{R^{x_3}\}\in {\cal {M}}_{C_0C; {\mathscr{X}}_3}$ for Alice, auxiliary states $\tau_{D_0}^{a_4}\in {\cal {H}}_{D_0}$ and ${\mathscr{X}}_4$-POVM $\{S^{x_4}\}\in {\cal{M}}_{DD_0; {\mathscr  {X}}_4}$ for Bob such that
\begin{align*}
\sum_{a_3,a_4,x_3,x_4}\beta_{a_3a_4}^{x_3x_4}P(x_3,x_4|a_3,a_4)>c_2,
\tag{A4}
\end{align*}
where $P(x_3,x_4|a_3,a_4)$ are joint conditional probability distributions computed as
\begin{align*}
P(x_3,x_4|a_3,a_4)=\textrm{Tr}[(R^{x_3}_{C_0C}\otimes S^{x_4}_{DD_0})(\tau^{a_3}_{D_0}\otimes \rho_{CD}\otimes \tau^{a_4}_{D_0})],
\tag{A5}
\end{align*}
$c_2$ denotes the maximal achievable classical bound of average gain in terms of the semiquantum game $\mathbbm{G}'_{sq}$, and ${\mathscr A}_i=\{a_i\}$. For all joint conditional probability distributions $P_c(x_3,x_4|a_3,a_4)$s derived from shared classical resources (hidden variable model \cite{Bell,PR}) or separable quantum states (hidden state model \cite{Cire}, see Figure S1(d)), we have
\begin{align*}
\sum_{a_3,a_4,x_3,x_4}\beta_{a_3a_4}^{x_3x_4}P_c(x_3,x_4|a_3,a_4)\leq c_2,
\tag{A6}
\end{align*}
where $P_c(x_3,x_4|a_3,a_4)$ can be any convex combination of independent distributions in the variables $x_3$ and $x_4$, i.e, $P_c(x_3,x_4|a_3,a_4)=\sum_{j}q_jP_j(x_3|a_3)P_j(x_4|a_4)$, $\{q_j\}$ is a probability distribution.

\begin{figure}
\begin{center}
\includegraphics[width=.85\textwidth]{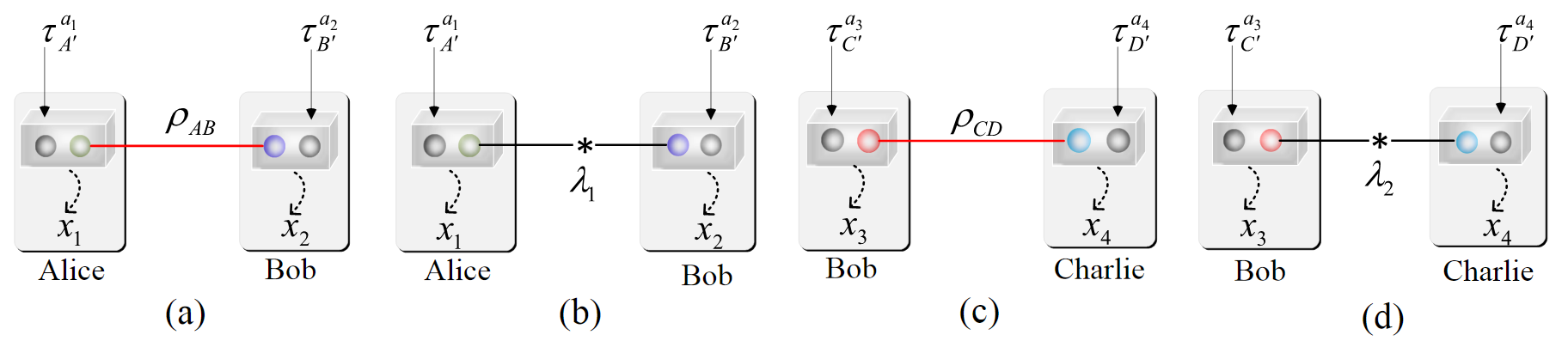}
\end{center}
\caption{(Color online) Schematically generalized Bell testing of bipartite entangled states derived from the semiquantum nonlocal game \cite{SQ}. (a) Generalized Bell nonlocality testing of one bipartite entanglement $\rho_{AB}$. $\tau^{a_1}_{A'}$ and $\tau^{a_2}_{B'}$ are input states of Alice and Bob, respectively, and can be sent from a trusty referee, where $A'$ and $B'$ are axillary systems, and $a_i\in {\mathscr{A}}_i$, $i=1, 2$. $x_1\in {\mathscr{X}}_1$ and $x_2\in {\mathscr {X}}_2$ are outputs (depending on special POVMs) of Alice and Bob, respectively. (b) Hidden state model for testing the locality of one shared source $\lambda_1$ or separable state $\varrho_{AB}=\sum_{i}p_i\varrho^{(i)}_{A}\otimes \varrho^{(i)}_{B}$, where $\{p_i\}$ is a probability distribution and $\varrho^{(i)}_{A(B)}$ are density operators of the system $A$ or $B$. (c) Generalized Bell nonlocality testing of one bipartite entanglement $\rho_{CD}$. $\tau^{a_3}_{C'}$ and $\tau^{a_4}_{D'}$ are input states of Bob and Charlie, respectively, and can be sent from a trusty referees, where $C'$ and $D'$ are axillary systems, and $a_i\in {\mathscr{A}}_i$, $i=3, 4$. $x_3\in {\mathscr{X}}_3$ and $x_4\in {\mathscr{X}}_4$ are outputs of Bob and Charlie, respectively. (d) Hidden state model for testing the locality of one shared source $\lambda_2$ or separable state $\varrho_{CD}=\sum_{i}q_i\varrho^{(i)}_{C}\otimes \varrho^{(i)}_{D}$, where $\{q_i\}$ is a probability distribution and $\varrho^{(i)}_{C(D)}$ are density operators of the system $C$ or $D$. The probabilities of $\tau^{a_1}, \cdots, \tau^{a_4}$ are uniform distributions in our assumptions.
}
\end{figure}

Note that the inequalities (A4) and (A6) can be regarded as generalized Bell-type inequalities for testing bipartite entanglement $\rho_{CD}$. In what follows, we construct a tripartite Bell inequality for testing the nonlocality of $\Lambda$-shaped quantum network consisting of $\rho_{AB}\otimes\rho_{CD}$ by taking use of the inequalities (A1) and (A4).

\begin{figure}
\begin{center}
\includegraphics[width=.8\textwidth]{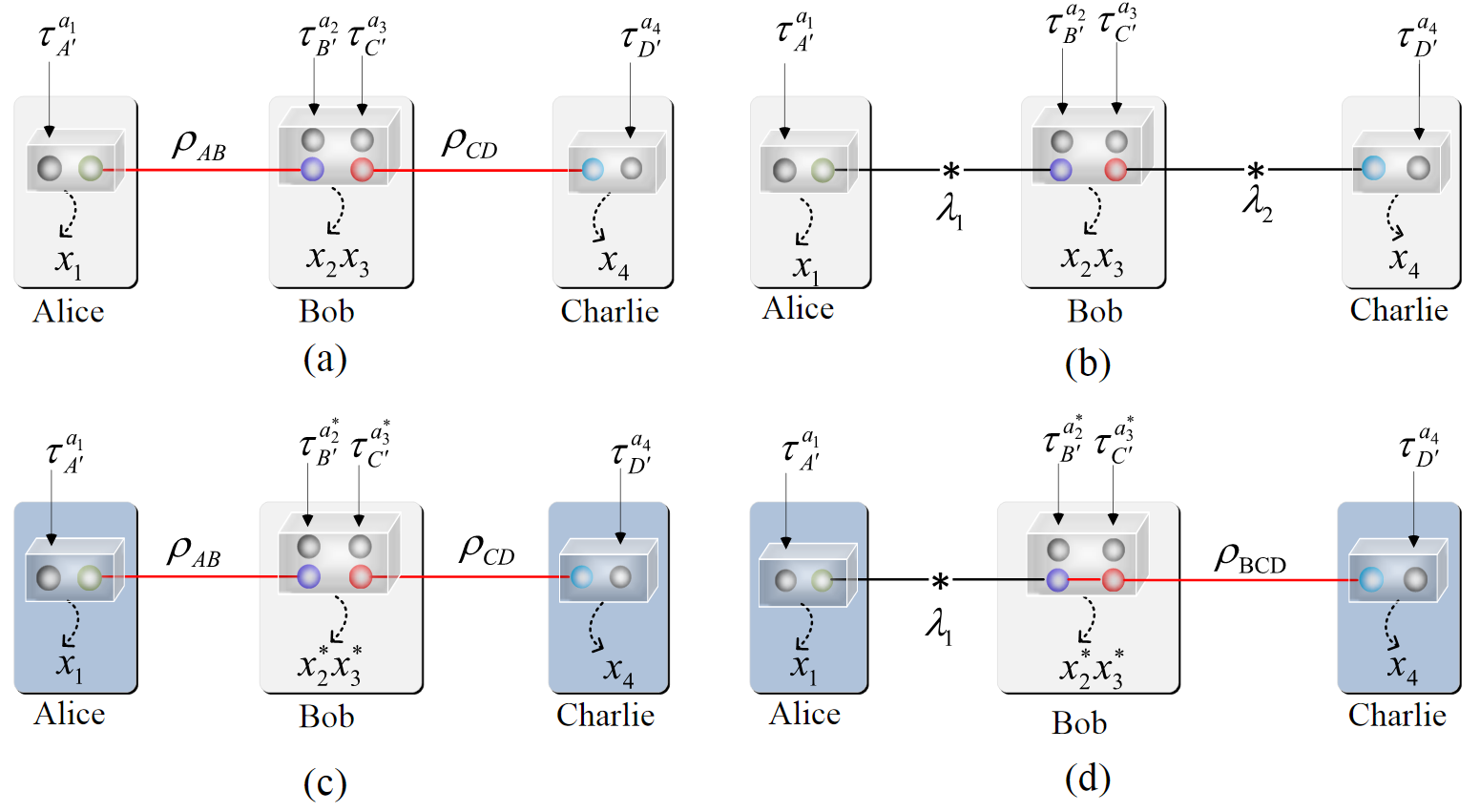}
\end{center}
\caption{(Color online) Schematically generalized Bell testing of $\Lambda$-shaped network. (a) Generalized nonlocality testing of a $\Lambda$-shaped quantum network consisting of two bipartite entangled states $\rho_{AB}$ and $\rho_{CD}$. Assume that Alice and Bob share the bipartite entanglement $\rho_{AB}$ while Bob and Charlie share the bipartite entanglement  $\rho_{CD}$. $\tau^{a_1}_{A_0}$, $\tau^{a_2}_{B_0}\otimes \tau^{a_3}_{C_0}$ and $\tau^{a_4}_{D_0}$ are input states of Alice, Bob, and Charlie, respectively, and can be sent from a trusty party, where $A_0, B_0, C_0, D_0$ are ancillary systems, and $a_i\in {\mathscr A}_i$, $i=1, \cdots, 4$. ${x}_1\in {\mathscr X}_1, {x}_2{x}_3\in {\mathscr X}_2\times {\mathscr X}_3$, and ${x}_4\in {\mathscr X}_4$ are outputs of Alice, Bob, and Charlie, respectively. (b) Hidden state model for testing the locality of a $\Lambda$-shaped network consisting of two shared sources $\lambda_1$ and $\lambda_2$ or separable states $\varrho_{ABCD}=\sum_{i}t_i\varrho^{(i)}_{a}\otimes \varrho^{(i)}_{BC}\otimes \varrho^{(i)}_{D}$, where $\{t_i\}$ is a probability distribution and $\varrho^{(i)}_{A(BC, D)}$ are density operators of the system $A$ (or $BC$, or $D$). Here, the joint system $BC$ can be entangled. (c) Generalized Bell testing of the activated nonlocality of the subnetwork consisting of Alice and Charlie. Bob firstly performs a proper POVM measurement $Q^{{x}^*_2{x}^*_3}_{B_0BCC_0}$ on his system and gets special output ${x}^*_2{x}^*_3$. And then, the standard bipartite Bell testing is performed by Alice and Bob for the collapsed state $\rho_{AD}=\textrm{Tr}_{B_0BCC_0}[Q^{{x}^*_2{x}^*_3}_{B_0BCC_0}(\tau^{a_2}_{B_0}\otimes \rho_{AB}\otimes \rho_{CD}\otimes \tau^{a_3}_{C_0})]$. (d) Hidden state model for testing the locality of the subnetwork consisting of Alice and Charlie, where Bob performs any POVM measurement before verifying the locality. The $\Lambda$-shaped network consists of two shared sources $\lambda_1, \lambda_2$, or one source $\lambda_1$ and a bipartite separable state entanglement $\varrho_{BCD}=$, or a tripartite fully separable state $\varrho_{A\,BC\,D}$). The probabilities of input states $\tau^{a_1}, \cdots, \tau^{a_4}$ are uniform distributions.
}
\end{figure}

\vskip 0.2cm
{\bf A1: Tripartite Bell testing experiment}
\vskip 0.2cm

Now, consider a new tripartite Bell testing experiment as shown in Figure S2(a) and Figure S2(b). Assume that the input and output sets of Bell testing experiments for $\rho_{AB}$ and $\rho_{CD}$ are ${\mathscr{A}}_i=\{1, 2, \cdots, m_i\}$, ${\mathscr{X}}_i=\{1, 2, \cdots, n_i\}$, respectively, $i=1, \cdots, 4$.

Assume that the input states of Alice, Bob and Charlie are $\tau^{{a}_1}_{A_0}$, $\tau^{{a}_2}_{B_0}\otimes \tau^{{a}_3}_{C_0}$ and $\tau^{{a}_4}_{D_0}$, respectively. Based on the POVMs $\{P^{a_1}\}, \{P^{a_2}\}$, $\{Q^{x_1}\},\{Q^{x_2}\}$, define new POVMs of three observers as $\{{P}^{{x}_1} \} \in {\cal M}_{A_0A; {\mathscr{X}}_1}$, $\{{Q}^{{x}_2{x}_3}\} \in {\cal M}_{B_0BCC_0; {\mathscr{X}}_2\times {\mathscr{X}}_3}$, and $\{{S}^{{x}_4} \}\in {\cal M}_{DD_0;{\mathscr{X}}_4}$, respectively.

From the equations (A2) and (A5), the joint conditional probability distribution $P({x}_1,{x}_2{x}_3, {x}_4|{a}_1, {a}_2{a}_3, {a}_4)$ is given as follows:
\begin{align*}
P({x}_1,{x}_2{x}_3,{x}_4|{a}_1, {a}_2{a}_3, {a}_4)=\textrm{Tr}[({P}^{{x}_1}\otimes {Q}^{{x}_2{x}_3}\otimes {S}^{{x}_4})(\tau^{{a}_1}_{A_0}\otimes \rho_{AB}\otimes \tau^{{a}_2}_{B_0}\otimes \tau^{{a}_3}_{C_0}\otimes \rho_{CD}\otimes \tau^{{a}_4}_{D_0}].
\tag{A7}
\end{align*}

Define an average gain $\wp$ depending on all conditional probability distributions $P({x}_1,{x}_2{x}_3,{x}_4|{a}_1, {a}_2{a}_3, {a}_4)$s as
\begin{align*}
\wp=\sum_{{a}_1, \cdots, {a}_4, \atop{{x}_1, \cdots, {x}_4}}\gamma^{{x}_1 \cdots {x}_4}_{{a}_1 \cdots {a}_4}P({x}_1, {x}_2{x}_3, {x}_4|{a}_1, {a}_2{a}_3, {a}_4),
\tag{A8}
\end{align*}
where the coefficients $\gamma^{{x}_1 \cdots {x}_4}_{{a}_1 \cdots {a}_4}$ satisfy $\gamma^{{x}_1 \cdots {x}_4}_{{a}_1 \cdots {a}_4}=\alpha^{{x}_1 {x}_2}_{{a}_1 {a}_2}+\beta^{{x}_3 {x}_4}_{{a}_3 {a}_4}$.

From the equations (A7) and (A8), consider the quantum tripartite correlations obtained by locally measuring the joint system of $\Lambda$-shaped quantum network (see Figure S2(a)) with local POVMs $\{{P}_{A_0A}^{{x}_1}\}\in {\cal M}_{A_0A; {\mathscr{X}}_1}$ for Alice, $\{{Q}_{B_0BCC_0}^{{x}_2{x}_3}\}\in {\cal M}_{B_0BCC_0; {\mathscr{X}}_2\times {\mathscr X}_3}$ for Bob, and $\{{S}_{DD_0}^{{x}_4}\}\in {\cal M}_{DD_0;{\mathscr{X}}_4}$ for Charlie. It is easy to obtain that
\begin{align*}
\wp=&\sum_{{a}_1, \cdots, {a}_4, \atop{{x}_1, \cdots, {x}_4}}\gamma^{{x}_1 \cdots {x}_4}_{{a}_1 \cdots {a}_4}P({x}_1,{x}_2{x}_3, {x}_4|{a}_1, {a}_2{a}_3, {a}_4)
\\
=&\sum_{{a}_1, \cdots, {a}_4, \atop{{x}_1, \cdots, {x}_4}}(\alpha^{x_1 x_2}_{{a}_1 a_2}+\beta^{x_3 x_4}_{{a}_3 a_4})P({x}_1,{x}_2|{a}_1, {a}_2)P({x}_3, {x}_4|{a}_3, {a}_4)
\\
=&
m_3m_4\sum_{{a}_1, {a}_2, {x}_1, {x}_2}\alpha^{x_1 x_2}_{{a}_1 a_2}P(x_1,x_2|a_1,a_2)
+m_1m_2\sum_{{a}_3, {a}_4, {x}_3, {x}_4}\beta^{x_3 x_4}_{{a}_3 a_4}P(x_3,x_4|a_3,a_4)
\tag{A9}\\
> &m_3m_4c_1+m_1m_2c_2,
\tag{A10}
\end{align*}
where the equation (A9) is from the equalities $\sum_{a_3, a_4, x_3, x_4}P(x_3,x_4|a_3,a_4)=\sum_{a_3, a_4}\sum_{x_3, x_4}P(x_3,x_4|a_3,a_4)=\sum_{a_3, a_4}1=m_3m_4$ (from the normalization of conditional probabilities) and $\sum_{a_1, a_2, x_1, x_2}P(x_1,x_2|a_1,a_2)=\sum_{a_1, a_2}\sum_{x_1, x_2}P(x_1,x_2|a_1,a_2)=\sum_{a_1, a_2}1=m_1m_2$. The inequality (A10) is from the inequalities (A1) and (A4).

In what follows, we estimate the upper bound of $\wp$ defined in the equation (A8) for classical tripartite correlations in terms of hidden state model \cite{Cire} as shown in Figure S2(b). In detail, from the inequalities (A3) and (A6), it is reasonable to define classical tripartite correlations $P_c({x}_1,{x}_2{x}_3,{x}_4|{a}_1, {a}_2{a}_3, {a}_4)$ as the joint conditional probabilities of measuring a  shared tripartite fully separable state $\varrho=\sum_{i}t_i\varrho^{(i)}_{A}\otimes \varrho^{(i)}_{BC}\otimes \varrho^{(i)}_{D}$ ($\{t_i\}$ is a probability distribution) with local POVMs $\{\hat{P}_{A_0A}^{{x}_1}\}\in {\cal{M}}_{A_0A; {\mathscr{X}}_1}, \{\hat{Q}_{B_0BCC_0}^{{x}_2{x}_3}\}\in {\cal{M}}_{B_0BCC_0; {\mathscr{X}}_2\times {\mathscr{X}}_3},  \{\hat{S}_{DD_0}^{{x}_4}\}\in {\cal{M}}_{DD_0; {\mathscr{X}}_4}$. We get an inequality from the equation (A10) as
\begin{align*}
\wp_c=&\sum_{{a}_1, \cdots, {a}_4, \atop{{x}_1, \cdots, {x}_4}}\gamma^{{x}_1\cdots {x}_4}_{{a}_1 \cdots {a}_4}P_c({x}_1, {x}_2{x}_3, {x}_4|{a}_1, {a}_2{a}_3, {a}_4)
\\
=&\sum_{{a}_1, \cdots, {a}_4, \atop{{x}_1, \cdots, \overline{x}_4}}\gamma^{{x}_1\cdots {x}_4}_{{a}_1 \cdots {a}_4}
\textrm{Tr}[(\hat{P}_{A_0A}^{{x}_1}\otimes\hat{Q}_{B_0BCC_0}^{{x}_2{x}_3}\otimes \hat{S}_{DD_0}^{{x}_4})
\\
&\times(\tau^{{a}_1}_{A_0}\otimes  \tau^{{a}_2}_{B_0}\otimes \tau^{{a}_3}_{C_0}\otimes \tau^{{a}_4}_{D_0}\otimes\sum_{i}t_i\varrho^{(i)}_{A}\otimes \varrho^{(i)}_{BC}\otimes \varrho^{(i)}_{D})]
\\
=&\sum_{{a}_1, \cdots, {a}_4, \atop{{x}_1, \cdots, {x}_4}}
(\alpha^{x_1 x_2}_{{a}_1 a_2}+\beta^{x_3 x_4}_{{a}_3 a_4})\textrm{Tr}[(\hat{P}_{A_0A}^{{x}_1}\otimes \hat{Q}_{B_0BCC_0}^{{x}_2{x}_3}\otimes \hat{S}_{DD_0}^{{x}_4})
\\
&\times(\tau^{{a}_1}_{A_0}\otimes  \tau^{{a}_2}_{B_0}\otimes \tau^{{a}_3}_{C_0}\otimes \tau^{{a}_4}_{D_0} \otimes \sum_{i}t_i\varrho^{(i)}_{A}\otimes \varrho^{(i)}_{BC}\otimes \varrho^{(i)}_{D})]
\\
=&\sum_{i}t_i\sum_{{a}_1, \cdots, {a}_4, \atop{{x}_1, \cdots, {x}_4}}(\alpha^{x_1 x_2}_{{a}_1 a_2}+\beta^{x_3 x_4}_{{a}_3 a_4})P^{(i)}(x_1|a_1)P^{(i)}(x_2x_3|a_2a_3)
P^{(i)}(x_4|a_4)
\tag{A11}
\\
=&\sum_{i}t_i[\sum_{{a}_1, \cdots, {a}_4, \atop{{x}_1, \cdots, {x}_4}}\alpha^{x_1 x_2}_{{a}_1 a_2}P^{(i)}(x_1|a_1)P^{(i)}(x_2,x_3|a_2,a_3)P^{(i)}(x_4|a_4)
\\
&+\sum_{{a}_1, \cdots, {a}_4, \atop{{x}_1, \cdots, {x}_4}}\beta^{x_3 x_4}_{{a}_3 a_4}P^{(i)}(x_1|a_1)P^{(i)}(x_2,x_3|a_2,a_3)P^{(i)}(x_4|a_4)]
\\
=&\sum_{i}t_i[\sum_{{a}_1, \cdots, {a}_4, \atop{{x}_1,  {x}_2}}\alpha^{x_1 x_2}_{{a}_1 a_2}P^{(i)}(x_1|a_1)P^{(i)}(x_2|a_2,a_3)
\\
&+\sum_{{a}_1, \cdots, {a}_4, \atop{{x}_3, {x}_4}}\beta^{x_3 x_4}_{{a}_3 a_4}P^{(i)}(x_3|a_2,a_3)P^{(i)}(x_4|a_4)]
\tag{A12}
\\
=&\sum_{i}t_i[\sum_{a_3,a_4}\sum_{{a}_1,{a}_2,\atop{x_1,  {x}_2}}\alpha^{x_1 x_2}_{{a}_1 a_2}P^{(i)}(x_1|a_1)P^{(i)}(x_2|a_2,a_3)
\\
&+\sum_{{a}_1, a_2}\sum_{a_3, a_4, \atop{x_3, x_4}}\beta^{x_3 x_4}_{{a}_3 a_4}P^{(i)}(x_3|a_2,a_3)P^{(i)}(x_4|a_4)]
\\
\leq&\sum_{i}t_i[\sum_{a_3,a_4}c_1
+\sum_{{a}_1, a_2}c_2]
\tag{A13}
\\
=&m_3m_4c_1+m_1m_2c_2.
\tag{A14}
\end{align*}
In the equation (A11), we have taken use of the following notations: $P^{(i)}(x_1|a_1)=\textrm{Tr}[\hat{P}_{A_0A}^{x_1}(\tau^{{a}_1}_{A_0}\otimes\varrho^{(i)}_{a})]$, $P^{(i)}(x_2x_3|a_2a_3)=\textrm{Tr}[\hat{Q}_{B_0BCC_0}^{x_2x_3} (\tau^{{a}_2}_{B_0}\otimes \tau^{{a}_3}_{C_0}\otimes  \varrho^{(i)}_{BC})]$, and $P^{(i)}(x_4|a_4)=\textrm{Tr}[\hat{S}_{DD_0}^{x_4}(\tau^{{a}_4}_{D_0} \otimes \varrho^{(i)}_{D})]$. In order to get the equation (A12) we have used the equalities $\sum_{x_3}P^{(i)}(x_2,x_3|a_2,a_3)=P^{(i)}(x_2|a_2,a_3)$, $\sum_{x_2}P^{(i)}(x_2,x_3|a_2,a_3)=P^{(i)}(x_3|a_2,a_3)$, $\sum_{x_1}P^{(i)}(x_1|a_1)=\sum_{x_4}P^{(i)}(x_4|a_4)=1$ for each $i,j$, $a_1, \cdots, a_4$. In inequality (A13), $P^{(i)}(x_1|a_1)P^{(i)}(x_2|a_2,a_3)$ are independent conditional probabilities in terms of the variables $x_1, x_2$; $P^{(i)}(x_3|a_2,a_3)P^{(i)}(x_4|a_4)$ are independent conditional probabilities in terms of the variables $x_3, x_4$. Hence, for each $i, j$, $a_2,a_3,a_4$, we can take use of the inequalities (A3) and (A6). The equation (A14) is from the fact that $\{t_{i}\}$ is a probability distribution.

Hence, the equation (A7) has defined a generalized Bell-type inequality for verifying the tripartite nonlocality of a $\Lambda$-shaped quantum network consisting of bipartite entangled states $\rho_{AB}\otimes\rho_{CD}$.

\vskip 0.2cm
{\bf A2: Activated nonlocality of $\Lambda$-shaped quantum network}
\vskip 0.2cm

To prove the activated nonlocality, it is sufficient to prove that there are local observables for all observers such that one local measurement of one observer can create one bipartite entanglement for other two observers. Note that Alice and Bob, or Bob and Charlie have shared one bipartite entanglement. It only needs to prove the activated locality of Alice and Charlie with the help of Bob. The proof of the activated nonlocality is completed by showing that the bipartite correlations of Alice and Charlie (after a local measurement of Bob) are inconsistent with these from any semiseparable state $\varrho_{ABCD}$ in terms of some generalized Bell-type inequality for two systems.

The proof is completed by two cases:

Case 1. $\Lambda$-shaped quantum network with qubit systems

In this case, we show that there is an entanglement $\rho_{AD}$ derived from $\rho_{AB}\otimes \rho_{CD}$ after performing a proper measurement on the joint system $BC$ by Bob. Note that all qubit-based bipartite mixed entangled states are distillable \cite{HHH}. It means that Alice and Bob can obtain one bipartite entangled pure state $|\Phi\rangle_{AB}$ from $\rho_{AB}^{\otimes n}$ by using local operations and classical communication (LOCC) when $n$ is large, where they do not need to obtain one maximally entangled pure state. Similarly, Bob and Charlie can obtain one bipartite entangled pure state $|\Psi\rangle_{CD}$ from $\rho_{AB}^{\otimes n}$ when $n$ is large by using LOCC. These quantum operations are reasonable because LOCC or local operations and shared randomness cannot create entanglement between two parties initially sharing no entanglement \cite{HHH,SQ}. For new quantum systems $ABCD$ in the state $|\Phi\rangle_{AB}\otimes|\Psi\rangle_{CD}$, it is easy to prove that the bipartite nonlocality of Alice and Charlie can be activated after a Bell measurement of Bob on his systems $BC$, i.e., the entanglement swapping holds for any bipartite entangled pure states \cite{GMT,Luo}. The activated nonlocality can be proved by using CHSH inequality \cite{CHSH,GMT,Luo}, where all bipartite entangled pure states can be verified using a universal Bell inequality-CHSH inequality.

\begin{figure}
\begin{center}
\includegraphics[width=.5\textwidth]{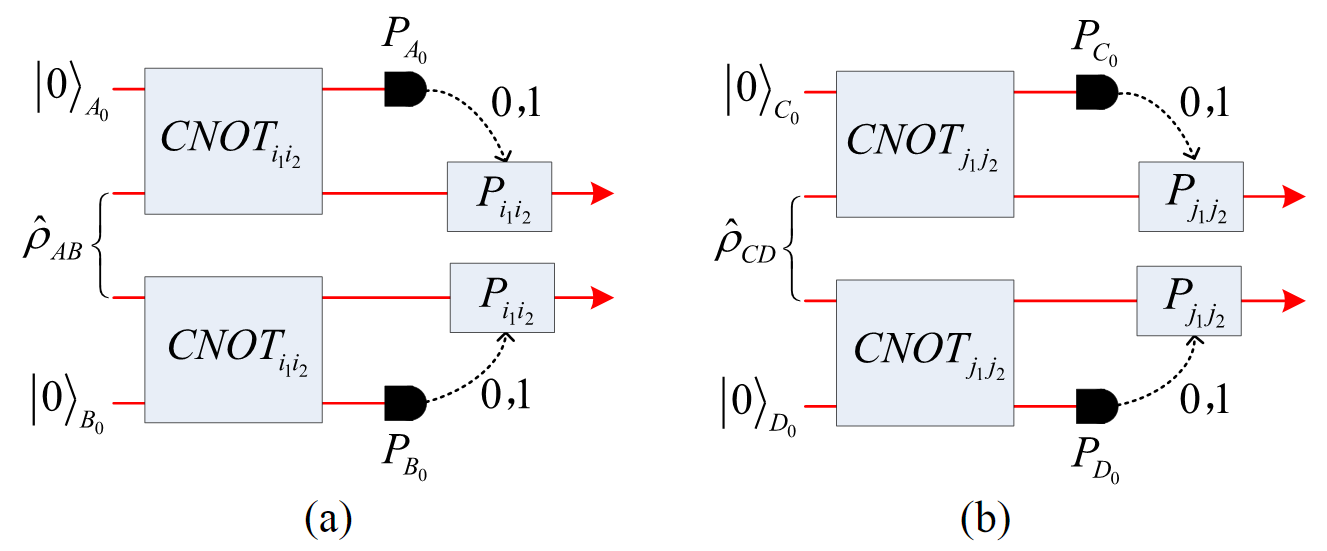}
\end{center}
\caption{(Color online) Schematic projection circuit of bipartite entangled states. (a) Project a bipartite entangled state $\hat{\rho}_{AB}$ into the subspace ${\cal H}_{i_1i_2}$ spanned by $\{|i_1i_1\rangle, |i_1i_2\rangle, |i_2i_1\rangle, |i_2i_2\rangle\}$. $\hat{\rho}_{AB}$ is shared by Alice and Bob. $CNOT_{i_1i_2}$ is controlled-NOT operation performed on the joint system of $A$ and $A_0$ by Alice, or $B$ and $B_0$ by Bob. It is defined as: $CNOT_{i_1i_2}: |i\rangle_{A}|0\rangle_{A_0}\mapsto |i\rangle_{A}|1\rangle_{A_0}$ for $i=i_1, i_2$,  and $CNOT_{i_1i_2}: |i\rangle_{A}|0\rangle_{A_0}\mapsto |i\rangle_{A}|0\rangle_{A_0}$ for all $i\not=i_1,i_2$; $CNOT_{i_1i_2}: |i\rangle_{B}|0\rangle_{B_0}\mapsto |i\rangle_{B}|1\rangle_{B_0}$ for $i=i_1, i_2$,  and $CNOT_{i_1i_2}: |i\rangle_{B}|0\rangle_{B_0}\mapsto |i\rangle_{B}|0\rangle_{B_0}$ for all $i\not=i_1,i_2$, where $A_0$ and $B_0$ are auxiliary systems in the state $|0\rangle$. $P_{A_0}$ or $P_{B_0}$ denotes the projection measurement under the basis $\{|0\rangle, |1\rangle\}$ on the system $A_0$ or $B_0$ respectively. $P_{i_1i_2}$ denotes the permutation operations: $|i_1\rangle\mapsto |0\rangle$ and $|i_2\rangle\mapsto |1\rangle$. $P_{i_1i_2}$ is performed if the measurement outcome is $1$. (b) Project a bipartite entangled state $\hat{\rho}_{CD}$ into the subspace ${\cal H}_{j_1j_2}$ spanned by $\{|j_1j_1\rangle, |j_1j_2\rangle, |j_2j_1\rangle, |j_2j_2\rangle\}$. $\hat{\rho}_{CD}$ is shared by Bob and Charlie. $CNOT_{j_1j_2}$ denotes controlled-NOT operation performed on the joint system of $C$ and $C_0$ by Bob, or $D$ and $D_0$ by Charlie. It is defined as: $CNOT_{j_1j_2}: |j\rangle_{C}|0\rangle_{C_0}\mapsto |j\rangle_{C}|1\rangle_{C_0}$ for $j=j_1, j_2$ and $CNOT_{j_1j_2}: |j\rangle_{C}|0\rangle_{C_0}\mapsto |j\rangle_{C}|1\rangle_{C_0}$ for $i=i_1, i_2$ all $j\not=j_1,j_2$;
 $CNOT_{j_1j_2}: |j\rangle_{D}|0\rangle_{D_0}\mapsto |j\rangle_{D}|1\rangle_{D_0}$ for $j=j_1, j_2$ and $CNOT_{j_1j_2}: |j\rangle_{D}|0\rangle_{D_0}\mapsto |j\rangle_{D}|1\rangle_{D_0}$ for $i=i_1, i_2$ all $j\not=j_1,j_2$, where $C_0$ and $D_0$ are auxiliary systems in the state $|0\rangle$. $P_{C_0}$ or $P_{D_0}$ denotes the projection measurement under the basis $\{|0\rangle, |1\rangle\}$ on the system $C_0$ or $D_0$ respectively. $P_{j_1j_2}$ denotes the permutation operations: $|j_1\rangle\mapsto |0\rangle$ and $|j_2\rangle\mapsto |1\rangle$. $P_{j_1j_2}$ is performed if the measurement outcome is $1$.}
\end{figure}

Case 2. $\Lambda$-shaped quantum network with high-dimensional systems

In this case, we show that there is a qubit-based entanglement $\rho_{AD}$ derived from high-dimensional entangled states $\rho_{AB}\otimes \rho_{CD}$ after performing proper projective measurements on the joint systems $BC$ by Bob. Assume that $\rho_{AB}=\sum_{i}p_i|\Phi_i\rangle_{AB}\langle \Phi_i|$ and $\rho_{CD}=\sum_{j}q_j|\Psi_j\rangle_{CD}\langle \Psi_j|$, where $|\Phi_i\rangle_{AB}, |\Psi_j\rangle_{CD}$ are entangled pure states for some $i,j$, $\{p_i\}$ and $\{q_j\}$ are probability distributions. To explain the main idea clearly, we only consider the simplest case in what follows, where similar proof can be easily followed for general cases. Assume that $\rho_{AB}$ and $\rho_{CD}$ has the following forms:
\begin{align*}
\rho_{AB}=p_1|\Phi_1\rangle_{AB}\langle \Phi_1|+p_2|\Phi_2\rangle_{AB}\langle \Phi_2|,
\tag{A15}
\\
\rho_{CD}=q_1|\Psi_1\rangle_{CD}\langle \Psi_1|+q_2|\Psi_2\rangle_{CD}\langle \Psi_2|,
\tag{A16}
\end{align*}
where $|\Phi_1\rangle_{AB}, |\Psi_1\rangle_{CD}$ are assumed to be bipartite entangled pure states. Let the bipartite decompositions of $|\Phi_1\rangle_{AB}, |\Psi_1\rangle_{CD}$ be  $|\Phi_1\rangle_{AB}=\sum_{i}r_i|\phi_i\rangle_A|{\phi}'_i\rangle_B, |\Psi_1\rangle_{CD}=\sum_{j}s_j|\psi_j\rangle_C|{\psi}'_j\rangle_D$, where $\{|\phi_i\rangle_A\}, \{|{\phi}'_i\rangle_B\}, \{|\psi_j\rangle_C\}, \{|\psi'_j\rangle_D\}$ are orthogonal basis states of the system $A, \cdots, D$, respectively, and $\{r_i\}$ and $\{s_j\}$ are probability distributions. After proper local unitary operations performed by three parties, i.e., $U: |\phi_i\rangle_A\mapsto |i\rangle_A$ by Alice, $V: |\phi'_i\rangle_B\mapsto |i\rangle_B$ and $W: |\psi_j\rangle_C\mapsto |j\rangle_C$ by Bob, and $Q:|\psi'_j\rangle_D\mapsto |j\rangle_D$ by Charlie, the systems of $\rho_{AB}, \rho_{CD}$ are changed as follows:
\begin{align*}
\hat{\rho}_{AB}&=(U\otimes V)\rho_{AB}(U^\dag\otimes V^\dag)
\\
=&p_1(\sum_{i}r_i|ii\rangle_{AB})(\sum_{i}r_i\langle ii|_{AB})+p_2|\hat{\Phi}_2\rangle_{AB}\langle \hat{\Phi}_2|,
\tag{A17}
\end{align*}
\begin{align*}
\hat{\rho}_{CD}&=(W\otimes Q)\rho_{AB}(W^\dag\otimes Q^\dag)
\\
=&q_1(\sum_{j}s_j|jj\rangle_{CD})(\sum_{j}r_j\langle jj|_{CD})
+q_2|\hat{\Psi}_2\rangle_{CD}\langle \hat{\Psi}_2|,
\tag{A18}
\end{align*}
where $|\hat{\Phi}_2\rangle_{AB}=(U\otimes V)|\Phi_2\rangle_{AB}$ and $|\hat{\Psi}_2\rangle_{CD}=(W\otimes Q)|\Psi_2\rangle_{CD}$.

Define ${\cal{H}}_{i_1i_2}$ as the subspaces spanned by $\{|i_1i_1\rangle, |i_1i_2\rangle, |i_2i_1\rangle, |i_2i_2\rangle\}$. By taking use of the quantum circuit (projective operations) shown in Figure S3(a), Alice and Bob can obtain one bipartite state $\varrho^{i_1i_2}_{AB}$ given by
\begin{align*}
\varrho^{i_1i_2}_{AB}=p_1(r_{i_1}|00\rangle+r_{i_2}|11\rangle)_{AB}(r_{i_1}\langle 00|+r_{i_2}\langle 11|)+q_1|\hat{\Phi}_{i_1i_2}\rangle_{AB}\langle \hat{\Phi}_{i_1i_2}|
\tag{A19}
\end{align*}
for the measurement outcome $00$ with a probability $p_{00}$, where $|\hat{\Phi}_{i_1i_2}\rangle_{AB}\in {\cal H}_{i_1i_2}$ can be a proper pure state. If $\varrho^{i_1i_2}_{AB}$ is entangled, Alice and Bob obtain a qubit-based bipartite entanglement. Otherwise, one of collapsed states of Alice and Bob for the measurement outcomes $01, 10 $ and $11$ is entangled because the linear superposition of these four states, i.e., $\hat{\rho}_{AB}$, is entangled, where we have taken use of the fact that the linear superposition of any separable states are separable. Hence, two observers can obtain a collapsed entanglement by performing projective operations and classical communications. And then they let $\varrho^{i_1i_2}_{AB}$ as the input of the quantum circuit shown in Figure S3(a). This procedure can be iteratively performed with LOCC and post-selections of Alice and Bob. Hence, Alice and Bob can obtain a qubit-based bipartite entanglement with nonzero success probability. Similarly, Bob and Charlie can probabilistically obtain a qubit-based entanglement $\varrho^{j_1j_2}_{CD}$ by iteratively performing LOCC and post-selection from the quantum circuit shown in Figure S3(b).

Similar to Case 1, if $\varrho^{i_1i_2}_{AB}$ and $\varrho^{j_1j_2}_{CD}$ are entangled pure states, the bipartite nonlocality of Alice and Charlie can be activated after a Bell measurement of Bob, using CHSH inequality \cite{CHSH,Gisin1,Luo}. Otherwise, they can obtain entangled pure states using the entanglement distillation \cite{HHH} if multiple copies of $\varrho^{i_1i_2}_{AB}$ and $\varrho^{j_1j_2}_{CD}$ are available. Therefore, we have completed verifying the activated nonlocality of Theorem 1. Another method is as follows: For a $\Lambda$-shaped network consisting of Alice, Bob and Charlie, they can obtain two bipartite entangled pure states. In this case, nonlinear Bell-type inequalities \cite{GMT,Luo} can be used to prove the $k$-partite nonlocality. $\blacksquare$

From Theorem 1, it is easy to verify the following generalized $\Lambda$-shaped quantum networks and chain-shaped quantum networks.

\begin{figure}
\begin{center}
\includegraphics[width=.85\textwidth]{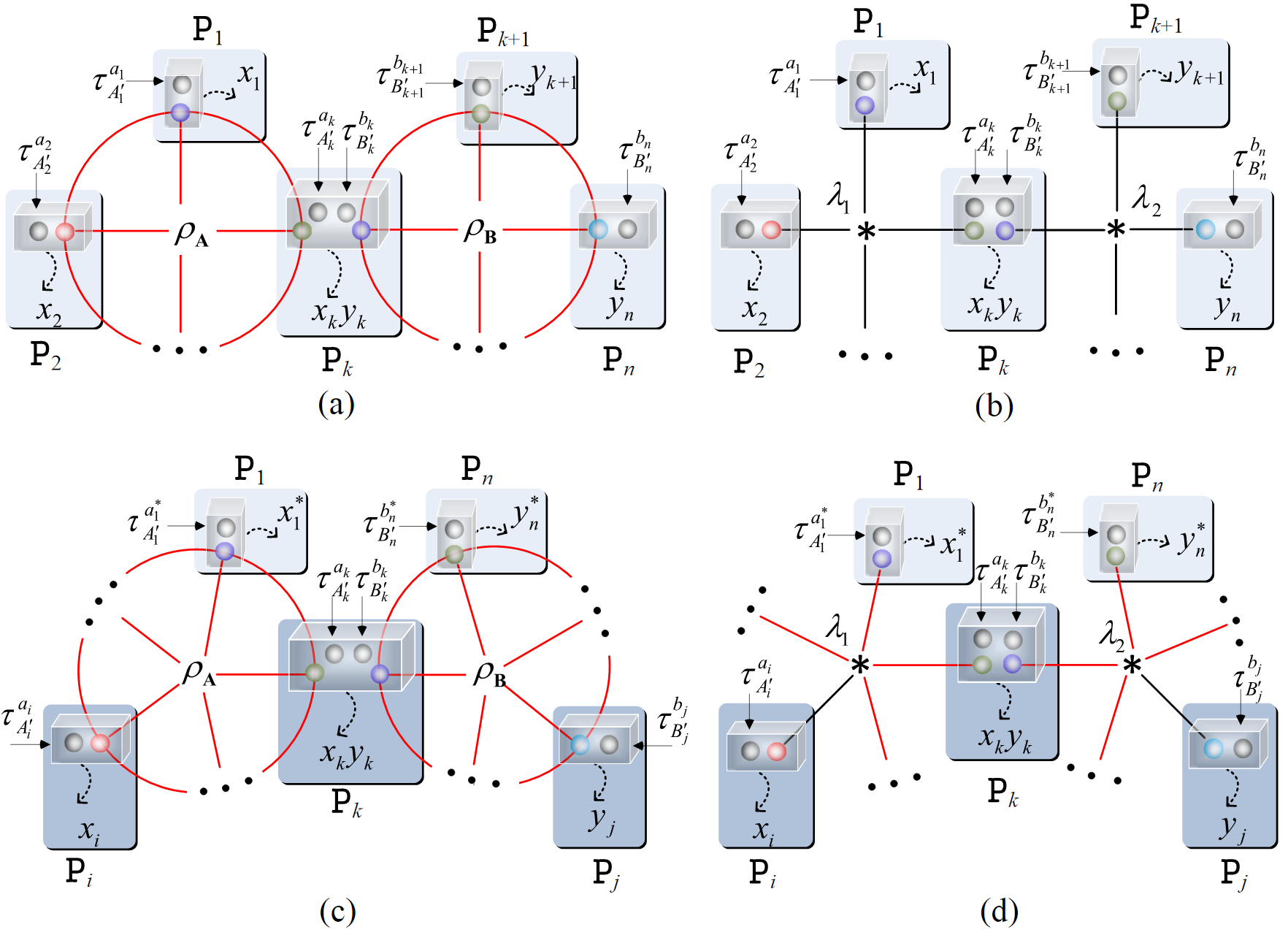}
\end{center}
\caption{(Color online) Schematically generalized Bell testing of a generalized $\Lambda$-shaped network. (a) Generalized nonlocality testing of a generalized $\Lambda$-shaped quantum network consisting of two entangled states $\rho_{\bf A}, \rho_{\bf B}$, where ${\textbf{A}}=A_{1}\cdots A_k$ and ${\bf B}=B_{k}\cdots B_n$. The observers $\texttt{P}_1$, $\cdots$, $\texttt{P}_k$ share one $k$-partite entanglement $\rho_{\textbf{A}}$ while the observers $\texttt{P}_k$, $\cdots$, $\texttt{P}_n$ share an $n-k+1$-partite entanglement $\rho_{\bf B}$. $\tau^{a_1}_{A'_1}$, $\tau^{a_2}_{A'_2}$, $\cdots$, $\tau^{a_k}_{A'_k}\otimes \tau^{b_k}_{B_k'}, \cdots, \tau^{b_n}_{B_n'}$ are input states of the observers $\texttt{P}_1$, $\cdots$, $\texttt{P}_k$, $\cdots$, $\texttt{P}_n$, respectively, and can be sent from a trusty referee, where $a_i\in {\mathscr A}_i$ and ${x}_i\in {\mathscr{X}}_i$ are respective inputs and outputs of the observer $\texttt{P}_i$, $i=1, \cdots, k$; $b_j\in {\mathscr{B}}_j$ and ${y}_j\in {\mathscr{Y}}_j$ are respective inputs and outputs of the observer $\texttt{P}_j$, $j=k, \cdots, n$. (b) Hidden state model for testing the locality of a generalized $\Lambda$-shaped network consisting of two shared sources $\lambda_1$ and $\lambda_2$ (or two bipartite separable states $\varrho_{\bf A}=\sum_{i}p_i\varrho^{(i)}_{A_1}\otimes \cdots \otimes \varrho^{(i)}_{A_k}$ and $\varrho_{\bf B}=\sum_{i}q_i\varrho^{(i)}_{B_k}\otimes \cdots \otimes \varrho^{(i)}_{B_n}$, where $\{p_i\}$ and $\{q_j\}$ are probability distributions and $\varrho^{(i)}_{A_j(B_s)}$ are density operators of the system $A_j$ (or $B_s$)). (c) Generalized Bell testing of the activated nonlocality of the subnetwork consisting of the observers $\texttt{P}_i$, $\texttt{P}_k$ and $\texttt{P}_j$ with a $\Lambda$-shaped system $\rho_{A_iA_k}\otimes\rho_{B_kB_j}$, where $\rho_{A_iA_k}$ is the collapsed state of $\rho_{\bf A}$ after a proper local measurement of all observers $\texttt{P}_1$, $\cdots$, $\texttt{P}_{k-1}$ except for the observer $\texttt{P}_i$; $\rho_{B_kB_n}$ is the collapsed state of $\rho_{\bf B}$ after a proper local measurement of all observers $\texttt{P}_{k+1}$, $\cdots$, $\texttt{P}_{n}$ except for the observer $\texttt{P}_j$, and $\tau^{a_s^*}, \tau^{b_t^*}$ and ${x}_s^*, {y}_t^*$ are respective special inputs and outputs. The new system is a standard $\Lambda$-shaped quantum network shown in Figure S2(a). (d) Hidden state model for testing the locality of the subnetwork consisting of three observers $\texttt{P}_i$, $\texttt{P}_k$ and $\texttt{P}_j$ with the reduced sources of $\lambda_1$ and $\lambda_2$ (or the collapsed states of any semi-separable states $\varrho_{\textbf{A}}=\sum_{s}p_s\varrho^{(s)}_{A_i}\otimes \varrho^{(s)}_{\hat{\textbf{A}}_i}$ and $\varrho_{\bf B}=\sum_{t}q_t\varrho^{(t)}_{B_j}\otimes \varrho^{(t)}_{\hat{\textbf{B}}_j}$, where $\{p_s\}$ and $\{q_t\}$ are probability distributions, and $\varrho^{(s)}_{\hat{\textbf{A}}_i}$ and $\varrho^{(t)}_{\hat{\bf B}_j}$ are density operators of the joint systems $\hat{\textbf{A}}_i=A_1\cdots A_{i-1}A_{i+1}\cdots A_k$ (or $\hat{\bf B}_j=B_k\cdots B_{j-1}B_{j+1}\cdots B_n$)), which can be entangled. The new system is a standard $\Lambda$-shaped network shown in Figure S2(b).}
\end{figure}

\textbf{Corollary 1}. For any generalized $\Lambda$-shaped quantum network ${\cal N}$ with $n$ observers $\texttt{P}_1$, $\texttt{P}_2$, $\cdots$, $\texttt{P}_n$, assume that $\texttt{P}_1, \cdots$, $\texttt{P}_k$ share one $k$-partite entanglement $\rho_{A_1\cdots A_k}$ on the Hilbert space ${\cal H}_{A_1}\otimes \cdots \otimes{\cal H}_{A_k}$, and $\texttt{P}_{k+1}, \cdots, \texttt{P}_n$ share an $n-k+1$-partite entanglement $\rho_{B_{k}\cdots B_n}$ on the Hilbert space ${\cal{H}}_{B_{k}}\otimes \cdots \otimes{\cal H}_{B_n}$. Then the joint system of $\rho_{A_1\cdots A_k}\otimes\rho_{B_{k}\cdots B_n}$ is $n$-partite nonlocal. Moreover, any two observers $\texttt{P}_i$ and $\texttt{P}_j$ with $1\leq i\leq k$, $k\leq j\leq n$ and $i\not=j$ has bipartite activated nonlocality after all the other observers in ${\cal{N}}$ perform proper local POVM.

{\bf Proof}. For a generalized $\Lambda$-shaped quantum network shown in Figure S4, from forward evaluations of Theorem 1 it is easy to prove the inconsistency of $n$-partite quantum correlations derived from the local POVMs of a generalized $\Lambda$-shaped quantum system $\rho_{A_1\cdots A_k}\otimes\rho_{B_{k}\cdots B_n}$ shown in Figure S4(a), and $n$-partite classical correlations from from the local measurements of any fully separable states $\varrho_{A_1\cdots A_k}\otimes\varrho_{B_{k}\cdots B_n}=(\sum_{i}p_i\varrho^{(i)}_{A_1}\otimes \cdots \otimes \varrho^{(i)}_{A_k})\otimes (\sum_{i}q_i\varrho^{(i)}_{B_k}\otimes \cdots \otimes \varrho^{(i)}_{B_n})$ shown in Figure S4(b)) in terms of the hidden state model \cite{Cire} using the semiquantum nonlocal game \cite{SQ} and similar procedure of the proofs given in Subsection A1 of Appendix A.

Now, consider the activated nonlocality of any quantum subnetwork. For any $k$-partite quantum entanglement $\rho_{A_1\cdots A_k}$, and two observers $\texttt{P}_i, \texttt{P}_j$ with $i,j\leq k$, there are local POVMs of all observers $\texttt{P}_s$ with $s\not=i,j$ and $s\leq k$ such that the collapsed state of two observers $\texttt{P}_i$ and $\texttt{P}_j$ after a local POVM of all the other observers $\texttt{P}_s$s is entangled. Similar result holds for the $n-k+1$-partite entanglement $\rho_{B_k\cdots B_n}$. So, for any three observers $\texttt{P}_i$, $\texttt{P}_k$ and $\texttt{P}_j$ with $1\leq i<k<j\leq n$, they can consist of a standard $\Lambda$-shaped quantum network after a proper local measurement of all the other observers as shown in Figure S4(c) and Figure S4(d). By using Theorem 1, it is easy to prove that the collapsed joint systems of three observers $\texttt{P}_i$, $\texttt{P}_k$ and $\texttt{P}_j$ have the bipartite activated nonlocality, i.e., the bipartite quantum correlations of $\texttt{P}_i$ and $\texttt{P}_k$ derived from a generalized $\Lambda$-shaped quantum network ${\cal N}$ after a local measurement of the observer $\texttt{P}_j$ (see Figure S4(c)) is consistent with all classical correlations from any separable systems (or further semi-separable states, i.e., three observers $\texttt{P}_i$, $\texttt{P}_k$ and $\texttt{P}_j$ have no prior shared entangled states while other observers can share some entangled states, see Figure S4(d)). $\blacksquare$

\begin{figure}
\begin{center}
\includegraphics[width=.85\textwidth]{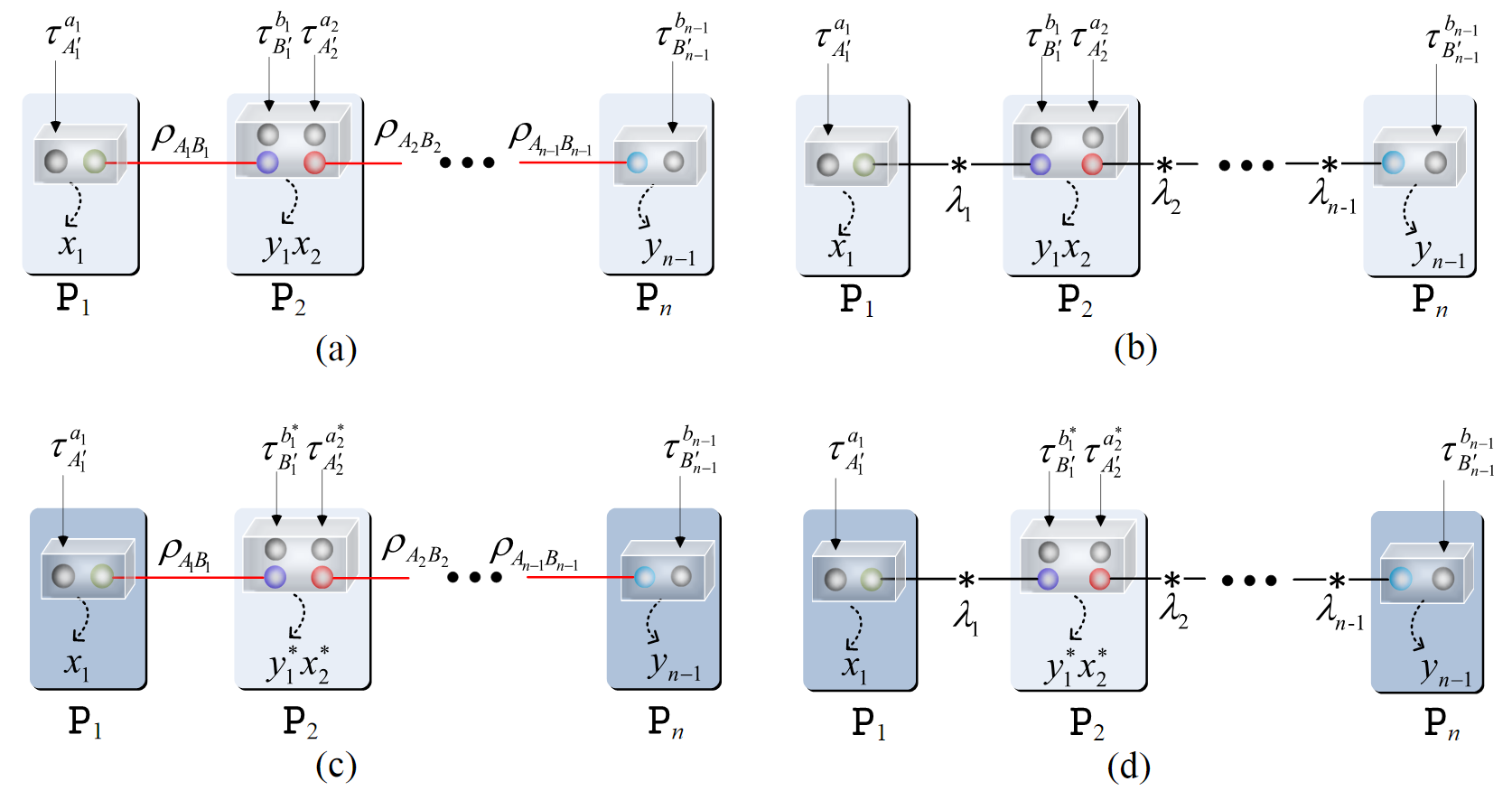}
\end{center}
\caption{(Color online) Schematically generalized Bell testing of a chain-shaped network. (a) Generalized nonlocality testing of a chain-shaped quantum network ${\cal N}_q$ consisting of $n-1$ bipartite entangled states $\rho_{A_1B_1}, \cdots, \rho_{A_{n-1}B_{n-1}}$, where the observers $\texttt{P}_i$ and $\texttt{P}_{i+1}$ share one bipartite entanglement $\rho_{A_iB_i}$, $i=1, \cdots, n-1$. $\tau^{a_1}_{A'_1}$ and ${x}_1$ are respective inputs and outputs of the observer $\texttt{P}_1$, $\tau^{b_i}_{B'_i}\otimes \tau^{a_{i+1}}_{A'_{i+1}}$ and ${y}_i{x}_{i+1}$ are respective inputs and outputs of the observer $\texttt{P}_i$ for $i=2, \cdots, i-2$, and $\tau^{b_{n-1}}_{B'_{n-1}}$ and ${y}_{n-1}$ are respective inputs and outputs of the observer $\texttt{P}_{n-1}$, where $A'_i$ and $B_i'$ are axillary systems that can be chosen by a trusty referee, $a_i\in {\mathscr{A}}_i, b_i\in {\mathscr{B}}_i, {x}_i\in {\mathscr{X}}_i, {y}_i\in {\mathscr{Y}}_i$. (b) Hidden state model for verifying the locality of a chain-shaped network consisting of $n-1$ shared sources $\lambda_1, \cdots, \lambda_{n-1}$ (or bipartite separable states $\varrho_{A_iB_i}=\sum_{s}p^{(i)}_s\varrho^{(s)}_{A_i}\otimes \varrho^{(s)}_{B_i}$, where $\{p^{(i)}_s\}$ are probability distributions and $\varrho^{(s)}_{A_i(B_i)}$ are density operators of the system $A_i$ or $B_i$, $i=1, \cdots, n-1$). (c) Generalized Bell testing of the activated nonlocality of the quantum subnetwork ${\cal N}_q^*\subset{\cal{N}}_q$ consisting of independent observers without initially sharing entangled states in ${\cal{N}}_q$. For example, ${\cal{N}}_q^*$ consists of two observers $\texttt{P}_1$ and $\texttt{P}_{n-1}$. Here, all observers $\texttt{P}_i\not\in{\cal N}^*$ perform a proper local POVM with special inputs $\tau^{a_i^*}, \tau^{b_i^*}$ and outputs ${x}_i^*, {y}_i^*$. (d) Hidden state model for verifying the activated locality of the subnetwork ${\cal{N}}^*$, where all parties $\texttt{P}_i\not\in{\cal{N}}^*$ can perform any POVM with any inputs $\tau^{a_i^*}, \tau^{b_i^*}$ and outputs ${x}_i^*,{y}_i^*$. The network consists of $n-1$ shared sources $\lambda_1, \cdots, \lambda_{n-1}$.}
\end{figure}

\textbf{Corollary 2}. For a chain-shaped quantum network ${\cal{N}}_q$ with $n$ observers $\texttt{P}_1$, $\texttt{P}_2$, $\cdots$, $\texttt{P}_n$, assume that two observers $\texttt{P}_i$ and $\texttt{P}_{i+1}$ share one bipartite entanglement $\rho_{A_iB_i}$ on the Hilbert space ${\cal{H}}_{A_i}\otimes {\cal{H}}_{B_i}$, $i=1, \cdots, n-1$. Then the joint system of $\rho_{A_1B_1}\otimes \cdots \otimes \rho_{A_{n-1}B_{n-1}}$ is $n$-partite nonlocal. Moreover, for any subnetwork ${\cal{N}}_q^*\subset{\cal{N}}_q$ consisting of independent observers (without initially sharing entangled states in ${\cal{N}}_q$), the reduced system of ${\cal{N}}_q^*$ is nonlocal after all observers who are not in the subnetwork of ${\cal{N}}_q^*$ perform a proper local POVM.

{\bf Proof}. For a chain-shaped quantum network shown in Figure S5, the $n$-partite nonlocality is followed from the semiquantum nonlocal game \cite{SQ} and the proof given in Subsection A1 of Appendix A, i.e., we can prove the inconsistency of $n$-partite quantum correlations derived from local measurements of a chain-shaped quantum network consisting of all bipartite entangled states $\rho_{A_1B_1}, \cdots, {\cal H}_{A_{n-1}B_{n-1}}$ shown in Figure S5(a), and $n$-partite classical correlations from a chain-shaped network consisting of fully separable states $\varrho_{A_1B_1}=\sum_{i}p^{(1)}_i\varrho^{(i)}_{A_1}\otimes \varrho^{(i)}_{B_1}$, $\cdots$, $\varrho_{A_{n-1}B_{n-1}}=\sum_{i}p^{(n-1)}_i\varrho^{(i)}_{A_{n-1}}\otimes \varrho^{(i)}_{B_{n-1}}$, as shown in Figure S5(b).

Now, consider the nonlocality of any quantum subnetwork consisting of independent observers without prior sharing entangled states. By using Theorem 1 iteratively it is easy to prove that each pair of independent observers $\texttt{P}_i$ and $\texttt{P}_j$ can share an entangled systems in the state $\rho_{A_iB_j}$ after all observers $\texttt{P}_k$s with $i<k<j$ performing a proper local POVM. Generally, consider a subnetwork ${\cal{N}}_q^*$ consisting of $k$ independent observers $\texttt{P}_{i_1}$, $\cdots$, $\texttt{P}_{i_k}$ with $1\leq i_1<\cdots <i_k\leq n$, where an example is shown in Figure S5(c) and Figure S5(d). For two observers $\texttt{P}_{i_j}$ and $\texttt{P}_{i_{j+1}}$, they can share a bipartite entanglement $\rho_{A_{i_j}B_{i_j}}$ after all observers $\texttt{P}_\ell$s with $i_j<\ell<i_{j+1}$ performing a proper local POVM. The activated bipartite nonlocality can be proved similar to the proof given in Subsection A2 of Appendix A (the proof of Theorem 1). Furthermore, by using the proof given in Subsection A1 of Appendix A, it is easy to prove that the joint systems of all independent observers $\texttt{P}_{i_1}$, $\cdots$, $\texttt{P}_{i_k}$ have $k$-partite nonlocality. Another method is as follows. For the subnetwork ${\cal{N}}_q^*$, all observers can obtain bipartite entangled pure states. In this case, the nonlinear Bell-type inequalities \cite{Luo} can be used to prove the $k$-partite nonlocality. $\blacksquare$

Additionally, the proof of Theorem 1 provided an interesting by-product that universal Bell inequality exists for detecting a single entanglement by using local projection and entanglement distilling \cite{HHH}. In fact, we have the following result

\textbf{Corollary 3}. There exists a universal Bell inequality to detect all entangled states with multiple copies.

{\bf Proof}. The proof is derived from the procedure shown in A2. For qubit-based entangled state $\rho_{A_1\cdots A_n}$ shared by parties $\texttt{P}_1, \cdots, \texttt{P}_1$, they can firstly perform a local entanglement distilling \cite{HHH} to obtain a multipartite entangled pure state $|\Phi\rangle_{A_1\cdots A_n}$. And then, $|\Phi\rangle_{A_1\cdots A_n}$ can be verified by using a universal Bell inequality \cite{YCZ}. For a high-dimensional entangled state $\rho_{A_1\cdots A_n}$, all parties can firstly perform a local projection shown in Figure S3 with LOCC in order to obtain a qubit-based entangled state $\hat{\rho}_{A_1\cdots A_n}$. Here, we have taken use of the fact that LOCC cannot create an entanglement among parties who have no initially shared entanglement. And then, all parties can perform a local entanglement distilling \cite{HHH} to obtain a multipartite entangled pure state $|\Psi\rangle_{A_1\cdots A_n}$ which can be verified by a universal Bell inequality \cite{YCZ}. The only assumption of the theorem is the multiple copies of single entanglement, which is reasonable in terms of statistics. $\blacksquare$

\section*{Appendix B: Proof of Theorem 2}

In this section, to complete the proof of Theorem 2 we firstly verify the nonlocality of all star-shaped quantum networks consisting of any entangled states.

\textbf{Lemma 1}. For a star-shaped quantum network ${\cal{N}}_q$ consisting of $n+1$ observers $\texttt{P}_1, \cdots, \texttt{P}_n$, and Bob, assume that two observers $\texttt{P}_i$ and Bob share one bipartite entanglement $\rho_{A_iB_i}$ on the Hilbert space ${\cal{H}}_{A_i}\otimes {\cal{H}}_{B_i}$, $i=1, \cdots, n$. Then the joint system of $A_1, B_1,\cdots, A_n, B_n$ is $n+1$-partite nonlocal. Moreover, there are local observables such that a local POVM of Bob can create an $n$-partite entanglement shared by $n$ observers $\texttt{P}_1, \cdots, \texttt{P}_n$ who are initially sharing no entangled states.

\textbf{Proof}. The proof is similar to that of Theorem 1. Since $\rho_{A_iB_i}$ is bipartite entangled, there is a semiquantum nonlocal game $\mathbbm{G}^{i}_{sq}$ \cite{SQ}, constants  $\alpha^{x_iy_i}_{a_ib_i}$, auxiliary states $\tau_{A'_i}^{a_i}\in {\cal{H}}_{A'_i}$ and ${\mathscr{X}}_i$-POVM $\{P^{x_i}\}\in {\cal{M}}_{A'_iA_i; {\mathscr{X}}_i}$ for $\texttt{P}_i$, auxiliary states $\tau_{B'_i}^{b_i}\in {\cal{H}}_{B'_i}$ and ${\mathscr{Y}}_i$-POVM $\{Q^{y_i}\}\in {\cal{M}}_{B_i'B_i; {\mathscr{Y}}_i}$ for Bob such that
\begin{align*}
\sum_{x_i,y_i,a_i,b_i}\alpha^{x_iy_i}_{a_ib_i}P(x_i,y_i|a_i,b_i)>c_i,
\tag{B1}
\end{align*}
where $P(x_i,y_i|a_i,b_i)$ are joint conditional probability distributions computed as
\begin{align*}
P(x_i,y_i|a_i,b_i)=\textrm{Tr}[(P^{x_i}_{A'_iA_i}\otimes Q^{y_i}_{B_i'B_i})(\tau^{a_i}_{A_i'}\otimes \rho_{A_iB_i}\otimes \tau^{b_i}_{B_i'})],
\tag{B2}
\end{align*}
$c_i$ denotes the maximal achievable classical bound of average gain in terms of the semiquantum nonlocal game $\mathbbm{G}^{i}_{sq}$, ${\mathscr{A}}_i=\{a_i\}$ and ${\mathscr{B}}_i=\{b_i\}$, ${\mathscr{X}}_i=\{x_i\}$ and ${\mathscr{Y}}_i=\{y_i\}$, $i=1, \cdots, n$. For all joint conditional probability distributions $P_c(x_i,y_i|a_i,b_i)$s derived from shared classical correlations or separable quantum states, we have
\begin{align*}
\sum_{x_i,y_i,a_i,b_i}\alpha^{x_iy_i}_{a_ib_i}P_c(x_i,y_i|a_i,b_i)\leq c_i,
\tag{B3}
\end{align*}
where $P_c(x_i,y_i|a_i,b_i)$ can be any convex combination of independent distributions in the variables $x_i,y_i$, i.e, $P_c(x_i,y_i|a_i,b_i)=\sum_{j}p_jP_j(x_i|a_i)P_j(y_i|b_i)$, $\{p_j\}$ is a probability distribution.

\begin{figure}
\begin{center}
\includegraphics[width=.9\textwidth]{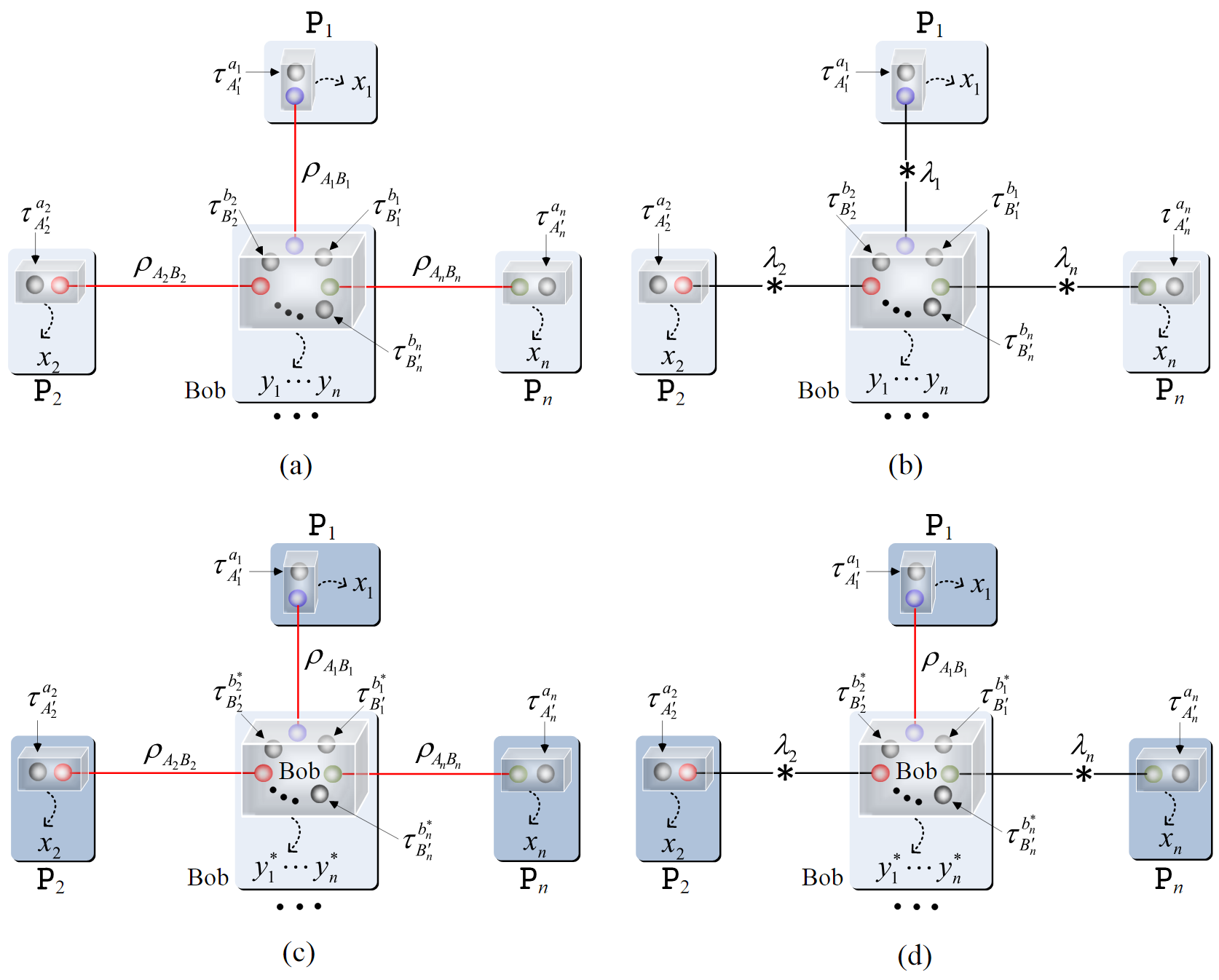}
\end{center}
\caption{(Color online) Schematically generalized Bell testing of a star-shaped network. (a) Generalized Bell nonlocality testing of a star-shaped quantum network consisting of $n$ bipartite entangled states $\rho_{A_1B_1}, \cdots, \rho_{A_nB_n}$, where two observers $\texttt{P}_i$ and Bob share one bipartite entanglement $\rho_{A_iB_i}$, $i=1, \cdots, n$. $\tau^{a_i}_{A'_i}$ and ${x}_i$ are respective inputs and outputs of the observer $\texttt{P}_i$, $\tau^{b_1}_{B'_1}\otimes \cdots \otimes \tau^{b_n}_{B'_n}$ and ${y}_1\cdots {y}_n$ are respective inputs and outputs of Bob, where $A'_i$ and $B_i'$ are axillary systems that can be chosen by a trusty referee, $a_i\in {\mathscr{A}}_i, b_i\in {\mathscr{B}}_i, {x}_i\in {\mathscr{X}}_i,{y}_i\in {\mathscr{Y}}_i$. (b) Hidden state model for verifying the locality of a star-shaped network consisting of $n$ shared sources $\lambda_1, \cdots, \lambda_n$ (or fully separable states $\varrho_{A_iB_i}=\sum_{s}p^{(i)}_s\varrho^{(s)}_{A_i}\otimes \varrho^{(s)}_{B_i}$, where $\{p^{(i)}_s\}$ are probability distributions and $\varrho^{(s)}_{A_i(B_i)}$ are density operators of the system $A_i$ or $B_i$, $i=1, \cdots, n$). (c) Generalized Bell testing of the activated nonlocality of the subnetwork consisting of all observers $\texttt{P}_1$, $\cdots$, $\texttt{P}_n$, where Bob performs a proper POVM with special inputs $\tau^{b_1^*}\otimes \cdots\otimes \tau^{b_n^*}$ and outputs $\overline{y}_1^*\cdots \overline{y}_n^*$ before verifying the nonlocality. (d) Hidden state model for verifying the locality of the subnetwork consisting of all observers $\texttt{P}_1$, $\cdots$, $\texttt{P}_n$, where Bob can perform any POVM with any inputs $\tau^{b_1^*}\otimes\cdots\otimes \tau^{b_n^*}$ and outputs $\overline{y}_1^* \cdots \overline{y}_n^*$. The network consists of $n$ shared sources $\lambda_1, \cdots, \lambda_n$, where one source ($\lambda_1$ for example) can be an entanglement.}
\end{figure}

In what follows, we construct an $n+1$-partite Bell inequality for testing the nonlocality of the star-shaped quantum network consisting of bipartite entangled states $\rho_{A_1B_1}\otimes\cdots \otimes \rho_{A_nB_n}$ from the inequalities (B1) and (B3).

\vskip 0.2cm
{\bf B1: Constructing $n+1$-partite generalized Bell testing}
\vskip 0.2cm

Now, we construct a new $n+1$-partite generalized Bell testing shown in Figure S6(a) and Figure S6(b). For each bipartite entanglement $\rho_{A_iB_i}$, assume that the input set and output set of the observer $\texttt{P}_i$ are ${\mathscr{A}}_i=\{1, 2, \cdots, m_i\}$, ${\mathscr{X}}_i=\{1, 2, \cdots, n_i\}$, respectively, and the input set and output set of Bob are ${\mathscr{B}}_i=\{1, 2, \cdots, \hat{m}_i\}$, ${\mathscr{Y}}_i=\{1, 2, \cdots, \hat{n}_i\}$, respectively, $i=1, \cdots, n$. Assume that the input states of the observers $\texttt{P}_1$, $\cdots$, $\texttt{P}_n$, and Bob are $\tau^{a_1}_{A_1'}$, $\cdots, \tau^{a_n}_{A_n'}$, $\tau^{b_1}_{B_1'}\otimes \cdots \otimes \tau^{b_n}_{B_n'}$, respectively. $P(\textbf{x}, \textbf{y}|\textbf{{a}}, \textbf{b})$ denote the joint conditional probability distributions defined by
\begin{align*}
P(\textbf{x}, \textbf{y}|\textbf{a}, \textbf{b})
=&\textrm{Tr}[({P}_{A_1'A_1}^{{x}_1}\otimes \cdots \otimes {P}_{A_n'A_n}^{{x}_n}\otimes  {Q}_{B_1\cdots B_nB_1'\cdots B_n'}^{{y}_1\cdots{y}_n})
(\otimes_{i=1}^n(\tau^{a_i}_{A_i'}\otimes \rho_{A_iB_i}\otimes\tau^{b_i}_{B_i'}))]
\\
=&\textrm{Tr}[({P}_{A_1'A_1}^{{x}_1}\otimes \cdots \otimes {P}_{A_n'A_n}^{{x}_n}\otimes  {Q}_{B_1'B_1}^{{y}_1}\otimes Q_{B_n'B_n}^{{y}_n})
(\otimes_{i=1}^n(\tau^{a_i}_{A_i'}\otimes \rho_{A_iB_i}\otimes\tau^{b_i}_{B_i'}))]
\\
=&\prod_{i=1}^nP(x_i,y_i|a_i,b_i),
\tag{B4}
\end{align*}
where
$P(x_i,y_i|a_i,b_i)=\textrm{Tr}[P^{{x}_i}_{A_i'A_i}\otimes Q^{y_i}_{B_i'B_i}) (\tau^{a_i}_{A_i'}\otimes \rho_{A_iB_i}\otimes\tau^{b_i}_{B_i'})]$, $\{{P}^{{x}_1}\}\in {\cal{M}}_{A_1'A_1; {\mathscr  X}_1}, \cdots, \{{P}^{{x}_n}\}\in {\cal M}_{A_n'A_n; {\mathscr{X}}_n}$, $\{{Q}^{{y}_1\cdots{y}_n}\}\in {\cal{M}}_{B_1'B_1\cdots B_n'B_n;{\mathscr{Y}}_1\times \cdots \times {\mathscr{Y}}_n}$ are POVMs of all observers $\texttt{P}_1$, $\cdots$, $\texttt{P}_n$, and Bob, respectively, $\textbf{x}=({x}_1, \cdots,{x}_n)$ ($n$ outputs of $n$ observers $\texttt{P}_1$, $\cdots$, $\texttt{P}_n$), $\textbf{y}={y}_1\cdots {y}_n$ (one output of Bob consisting of $n$ indexes), ${\textbf{a}}=(a_1, \cdots, a_n)$ (the indexes of $n$ inputs of $n$ observers $\texttt{P}_1$, $\cdots$, $\texttt{P}_n$), $\textbf{b}=b_1\cdots b_n$ (the index of one input of Bob consisting of $n$ indexes).

Define the average gain $\wp$ depending on all joint conditional probabilities $P(\textbf{x}, \textbf{y}|\textbf{a},\textbf{b})$s as
\begin{align*}
\wp=\sum_{\textbf{{x}}, \textbf{{y}},\textbf{a}, \textbf{b}}
\gamma^{\textbf{x} \textbf{y}}_{\textbf{a} \textbf{b}}
P(\textbf{x}, \textbf{y}|\textbf{a},\textbf{b}),
\tag{B5}
\end{align*}
where the coefficients $\gamma^{\textbf{x}\textbf{y}}_{\textbf{a} \textbf{b}}$ are given by $\gamma^{\textbf{x}\textbf{y}}_{\textbf{a} \textbf{b}}=\sum_{i=1}^n\alpha^{{x}_i {y}_i}_{a_i b_i}$.

From the equations (B4) and (B5), consider the quantum $n+1$-partite correlations obtained from the locally measuring the shared system of a star-shaped quantum network shown in Figure S6(a) with POVMs $\{{P}_{A_1'A_1}^{{x}_1}\}\in {\cal M}_{A_1'A_1; {\mathscr{X}}_1}$ for the observer $\texttt{P}_1$, $\cdots$,  $\{{P}_{A_n'A_n}^{{x}_n}\}\in {\cal{M}}_{A_n'A_n; {\mathscr  X}_n}$ for the observer $\texttt{P}_n$, and $\{{Q}_{B_1'B_1\cdots B_n'B_n}^{{y}_1\cdots{y}_n}\}\in {\cal{M}}_{B_1'B_1\cdots B_n'B_n; {\mathscr{Y}}_1\times\cdots \times {\mathscr{Y}}_n}$ for Bob. The equation (B5) is rewritten into

\begin{align*}
\wp=&
\sum_{\textbf{a},\textbf{b},
\textbf{x},\textbf{y}}\gamma^{\textbf{x} \textbf{y}}_{\textbf{a} \textbf{b}}
P(\textbf{x},\textbf{y}|\textbf{a}, \textbf{b})
\\
=&\sum_{\textbf{a},\textbf{b},
\textbf{x},\textbf{y}}
(\alpha^{x_1 y_1}_{a_1 b_1}+\cdots +\alpha^{x_n y_n}_{a_n b_n})
P(\textbf{x}, \textbf{y}|\textbf{a}, \textbf{b})
\\
=&\sum_{\textbf{a},\textbf{b},
\textbf{x},\textbf{y}}
(\alpha^{x_1 y_1}_{a_1 b_1}+\cdots +\alpha^{x_n y_n}_{a_n b_n})
\prod_{i=1}^nP(x_i,y_i|a_i,b_i)
\\
=&\sum_{i=1}^n\frac{\hat{M}}
{{m}_i\hat{m}_i}\sum_{a_i,b_i,x_i,y_i}\alpha^{x_i y_i}_{a_i b_i}
P(x_i,y_i|a_i,b_i)
\tag{B6}
\\
> &
\sum_{i=1}^n\frac{\hat{M}c_i}{{m}_i\hat{m}_i}
\tag{B7}
\end{align*}
Here, the equation (B6) is from the equalities $\sum_{{a}_j,{b}_j,
{x}_j,{y}_j}P(x_j,y_j|a_j,b_j)=m_j\hat{m}_j$ from the normalization of conditional probabilities, i.e., $\sum_{x_s,y_s}P(x_s,y_s|a_s,b_s)=1$ for each $a_s,b_s$, and  $\hat{M}=\prod_{i=1}^nm_i\hat{m}_i$. The inequality (B7) is from the inequalities (B1).

In what follows, we estimate the upper bound of $\wp$ defined in the equation (B5) for classical $n+1$-partite correlations by using hidden state model \cite{Cire}, see Figure S6(b). In detail, from the inequalities (B2), define classical $n+1$-partite correlations $P_c(\textbf{x}, \textbf{y}|\textbf{a}, \textbf{b})$ as the conditional probabilities of locally measuring a shared fully separable state $\varrho=\sum_{i}t_i\varrho^{(i)}_{A_1}\otimes \cdots \otimes \varrho^{(i)}_{A_n}\otimes \varrho^{(i)}_{B_1\cdots B_n}$ ($\{t_i\}$ is a probability distribution) with POVMs $\{\hat{P}_{A_1A_1'}^{{x}_1}\}\in {\cal{M}}_{A_1'A_1; {\mathscr{X}}_1}, \cdots, \{\hat{P}_{A_nA_n'}^{{x}_n}\}\in {\cal{M}}_{A_n'A_n; {\mathscr{X}}_n}$, $\{\hat{Q}_{B_1'B_1\cdots B_n'B_n}^{{y}_1\cdots{y}_n}\}\in {\cal M}_{B_1'B_1\cdots B_n'B_n; {\mathscr{Y}}_1\times \cdots\times {\mathscr{Y}}_n}$. We get an inequality from the equation (B5) as
\begin{align*}
\wp_c=&
\sum_{\textbf{a},\textbf{b},\textbf{x},\textbf{y}}
\gamma^{\textbf{x}\textbf{y}}_{\textbf{a} \textbf{b}}
P_c(\textbf{x},\textbf{y}|\textbf{a},\textbf{b})
\\
=&\sum_{\textbf{{a}},\textbf{{b}},
\textbf{{x}},\textbf{{y}}}
\textrm{Tr}[(\hat{P}_{A_1'A_1}^{{x}_1}\otimes \cdots \otimes \hat{P}_{A_n'A_n}^{{x}_n}\otimes \hat{Q}_{B_1'B_1\cdots B_n'B_n}^{{y}_1\cdots {y}_n})
\\
&\times(\tau^{a_1}_{A_1'}\otimes\cdots \otimes  \tau^{a_n}_{A_n'}\otimes \tau^{b_1}_{B_1'}\otimes \cdots
\otimes \tau^{b_n}_{B_n'}
\otimes\sum_{i}t_i\varrho^{(i)}_{A_1}\otimes \cdots \otimes \varrho^{(i)}_{A_n}\otimes \varrho^{(i)}_{B_1\cdots B_n}]
\\
=&\sum_{i}t_i\sum_{\textbf{a},\textbf{b},\textbf{x},\textbf{y}}
\sum_{j=1}^n\alpha^{x_j y_j}_{a_j b_j}\textrm{Tr}[(\hat{P}_{A_1'A_1}^{x_1}\otimes \cdots \otimes \hat{P}_{A_n'A_n}^{x_n}\otimes \hat{Q}_{B_1'B_1\cdots B_n'B_n}^{y_1\cdots{y}_n})
\\
&\times(\tau^{a_1}_{A_1'}\otimes\cdots \otimes  \tau^{a_n}_{A_n'}\otimes \tau^{b_1}_{B_1'}\otimes \cdots
\otimes \tau^{b_n}_{B_n'}
\otimes\varrho^{(i)}_{A_1}\otimes \cdots \otimes \varrho^{(i)}_{A_n}\otimes \varrho^{(i)}_{B_1\cdots B_n}]
\\
=&\sum_{i}t_i\sum_{\textbf{a},\textbf{b},\textbf{x},\textbf{y}}
\sum_{j=1}^n\alpha^{x_j y_j}_{a_j b_j}\prod_{s=1}^nP^{(i)}({x}_s|a_s)
P^{(i)}({y}_1\cdots{y}_n|b_1\cdots b_n)
\tag{B8}
\\
=&\sum_{i}t_i\sum_{j=1}^n\sum_{\textbf{a},\textbf{b}, x_j,y_j}\alpha^{x_j y_j}_{a_j b_j}P_k^{(i)}(x_j|a_j)
P^{(i)}(y_j|b_1\cdots b_n)
\tag{B9}
\\
=&\sum_{i}t_i\sum_{j=1}^n
\sum_{a_s,b_s, \atop{1\leq s\not=j\leq n}}\sum_{a_j,b_j,x_j,y_j}\alpha^{x_j y_j}_{a_j b_j}P^{(i)}(x_j|a_j)P^{(i)}(y_j|b_1\cdots b_n)
\\
\leq &\sum_{i}t_i\sum_{j=1}^n
\sum_{a_s,b_s,\atop{
1\leq s\not=j\leq n}}c_j
\tag{B10}
\\
=&\sum_{i=1}^n\frac{\hat{M}c_i}{m_i\hat{m}_i}.
\tag{B11}
\end{align*}
In the equation (B8), we have taken use of the following notations: $P^{(i)}({x}_j|a_j)=\textrm{Tr}[\hat{P}_{A_j'A_j}^{{x}_j}(\tau^{a_j}_{A_j'}\otimes\varrho^{(i)}_{A_j})]$, $P^{(i)}({y}_1\cdots{y}_n|b_1\cdots b_n)=\textrm{Tr}[\hat{Q}_{B_1'B_1\cdots B_n'B_n}^{{y}_1\cdots{y}_n} (\tau^{b_1}_{B_1'}\otimes \cdots \otimes \tau^{b_n}_{B_n'}\otimes  \varrho^{(i)}_{B_1\cdots B_n})]$, $j=1, \cdots, n$.  In order to get the equation (B9) we have used the equalities $\sum_{y_t, 1\leq t\not=j\leq n}P^{(i)}(y_1\cdots y_n|b_1\cdots b_n)=P^{(i)}(y_j|b_1\cdots b_n)$ and $\sum_{x_j}P^{(i)}(x_j|a_j)=1$ for each $i,j,a_1,\cdots, a_n, b_1, \cdots, b_n$. To get the inequality (B10), note that $P^{(i)}(x_j|a_j)$ and $P^{(i)}(y_j|b_1\cdots b_n)$ are independent conditional probabilities in terms of the variables $x_j, y_j$ for any fixed variables $a_j, b_j$. Hence, for each $i, j, k$, $x_j,y_j$, we can take use of the inequalities (B3). The equation (B11) is from the fact that $\{{t}_{i}\}$ is a probability distribution.

From the inequalities (B7) and (B11), we have verify the $n+1$-partite nonlocality of the star-shaped quantum network. Here, the inequalities (B7) and (B11) have defined $n+1$-partite generalized Bell-type inequalities for verifying the $n+1$-partite nonlocality of a star-shaped joint system $\rho_{A_1B_1}\otimes \cdots \otimes \rho_{A_nB_n}$.

Up to now, we have proved that the inequalities (B7) and (B11) can ensure that the $n+1$-partite quantum correlations derived from a star-shaped quantum network with entangled states are different from the $n+1$-partite classical correlations derived from the fully separable state.

\vskip 0.2cm
{\bf B2: Activated nonlocality of a star-shaped quantum network}
\vskip 0.2cm

To complete the proof, we further prove that the $n$-partite nonlocality of all independent observers, i.e., $\texttt{P}_1$, $\cdots$, $\texttt{P}_n$, can be activated by a local measurement of Bob, see Figure S6(c) and Figure S6(d). From the proof in subsection A2 of Appendix A, two observers $\texttt{P}_i$ and Bob can probabilistically obtain a qubit-based entangled state from their shared state $\rho_{A_iB_i}$. And then, by using the entanglement distill \cite{HHH} with LOCC, $\texttt{P}_i$ and Bob can share an entangled pure state $|\Phi\rangle_{A_iB_i}$, where LOCC cannot create an entanglement between two parties who have not initially shared entanglement.

In what follows, we only need to prove that the $n$-partite nonlocality of $n$ observers $\texttt{P}_1$, $\cdots$, $\texttt{P}_n$, can be activated by a local measurement of Bob for any star-shaped quantum network consisting of all entangled pure states $\otimes_{i=1}^n|\Phi\rangle_{A_iB_i}$. In detail, assume $|\Phi\rangle_{A_iB_i}=u_{i}|00\rangle+v_i|11\rangle$ under some local operations, where $u_{i}^2+v_i^2=1$, $i=1, \cdots, n$. Define the measurement of Bob as $n$-particle Bell basis, i.e., $\{Q_\pm^{y_1\cdots y_n}=\frac{1}{\sqrt{2}}(|y_1\cdots y_n\rangle\pm|\overline{y}_1\cdots \overline{y}_n\rangle)\}$, where $\overline{y}_i=1\oplus y_i$. For each measurement $Q_\pm^{y_1\cdots y_n}$, the collapsed state of all observers $\texttt{P}_1$, $\cdots$, $\texttt{P}_n$ is given by
\begin{align*}
\Omega^{\pm}_{A_1\cdots A_n}=r(u_{y_1\cdots y_n}|y_1\cdots y_n\rangle\pm v_{y_1\cdots y_n} |\overline{y}_1\cdots \overline{y}_n\rangle)
\tag{B12}
\end{align*}
which is an $n$-partite generalized GHZ state, where $u_{y_1\cdots y_n}=\prod_{i=1}^nu_{y_i}$, $v_{y_1\cdots y_n}=\prod_{i=1}^nv_{y_i}$ and $r$ is the normalization constant.

For any $n\geq 2$, we can easily prove the following $n$-partite CHSH inequality
\begin{align*}
&|\langle {M}^{1}_0\otimes \cdots \otimes{M}^{n-1}_0\otimes {M}^n_0\rangle
+\langle{M}^{1}_0\otimes\cdots \otimes {M}^{n-1}_0\otimes {M}^n_1\rangle
\nonumber\\
&+\langle{M}^{1}_1\otimes\cdots \otimes{M}^{n-1}_1\otimes{M}^n_0\rangle
-\langle{M}^{1}_1\otimes\cdots \otimes{M}^{n-1}_1\otimes{M}^n_1\rangle| \leq 2,
\tag{B13}
\end{align*}
where one can schematically represent ${M}_i:={M}^{1}_i\otimes\cdots \otimes{M}^{n-1}_i$ ($i=0, 1$) and prove the inequality by a similar procedure of CHSH inequality \cite{CHSH}. This $n$-partite Bell inequality is used to prove that $\Omega^{\pm}_{A_1\cdots A_n}$ are entangled for all nonzero $u_i,v_i$.
\begin{itemize}
\item{} For an even $m$, define dichotomic observables $M^j_i=(1-i)\sigma_z+i\sigma_x$, $i=0, 1; j=1, \cdots, n-1$, and $M^n_i=\cos\theta\sigma_z+(-1)^i\sin\theta \sigma_x$ with $\cos\theta=1/\sqrt{1+4r^2u_{y_1\cdots y_n}^2v^2_{y_1\cdots y_n}}$, where $\sigma_z$ and $\sigma_x$ are Pauli matrices. From straight forward evaluations we get
\begin{align*}
&\langle {M}^{1}_0\otimes \cdots \otimes{M}^{n-1}_0\otimes {M}^n_0\rangle
+\langle{M}^{1}_0\otimes\cdots \otimes {M}^{n-1}_0\otimes {M}^n_1\rangle
\nonumber\\
&+\langle{M}^{1}_1\otimes\cdots \otimes{M}^{n-1}_1\otimes{M}^n_0\rangle
-\langle{M}^{1}_1\otimes\cdots \otimes{M}^{n-1}_1\otimes{M}^n_1\rangle
\nonumber\\
=&{\rm tr}(({M}^{1}_0\otimes \cdots \otimes{M}^{m-1}_0\otimes {M}^n_0
+{M}^{1}_0\otimes\cdots \otimes {M}^{n-1}_0\otimes {M}^n_1
\nonumber\\
&+{M}^{1}_1\otimes\cdots \otimes{M}^{n-1}_1\otimes{M}^n_0
-{M}^{1}_1\otimes\cdots \otimes{M}^{n-1}_1\otimes{M}^n_1)\Omega^{\pm}_{A_1\cdots A_n}(\Omega^{\pm}_{A_1\cdots A_n})^\dag)
\nonumber\\
=&2\sqrt{1+4r^2u_{y_1\cdots y_n}^2v^2_{y_1\cdots y_n}}
\nonumber\\
>&2,
\tag{B14}
\end{align*}
which violates the multipartite CHSH inequality shown in equation (B13) for all nonzero $u_i,v_i, i=1, \cdots, n$. It implies that $\Omega^{\pm}_{A_1\cdots A_n}$ are $n$-partite nonlocal.

\item{} For an odd $m$, by defining observables $M^j_i=(1-i)\sigma_z+i\sigma_x$, $i=0, 1; j=1, \cdots, n-2$, $M^{n-1}_i=(1-i){\bf I}_2+i\sigma_x$, and $M^n_i=\cos\theta\sigma_z+(-1)^i\sin\theta \sigma_x$ with $\cos\theta=1/\sqrt{1+(1-v)^2}$, we obtain the same inequality (B14) which implies that $\Omega^{\pm}_{A_1\cdots A_n}$ violates the Bell inequality given in equation (B13), where ${\bf I}_2$ is the identity matrix.
\end{itemize}

Consequently, we have proved the lemma 1. $\blacksquare$

The Lemma 1 can be easily extended to generalized star-type networks.

\begin{figure}
\begin{center}
\includegraphics[width=.75\textwidth]{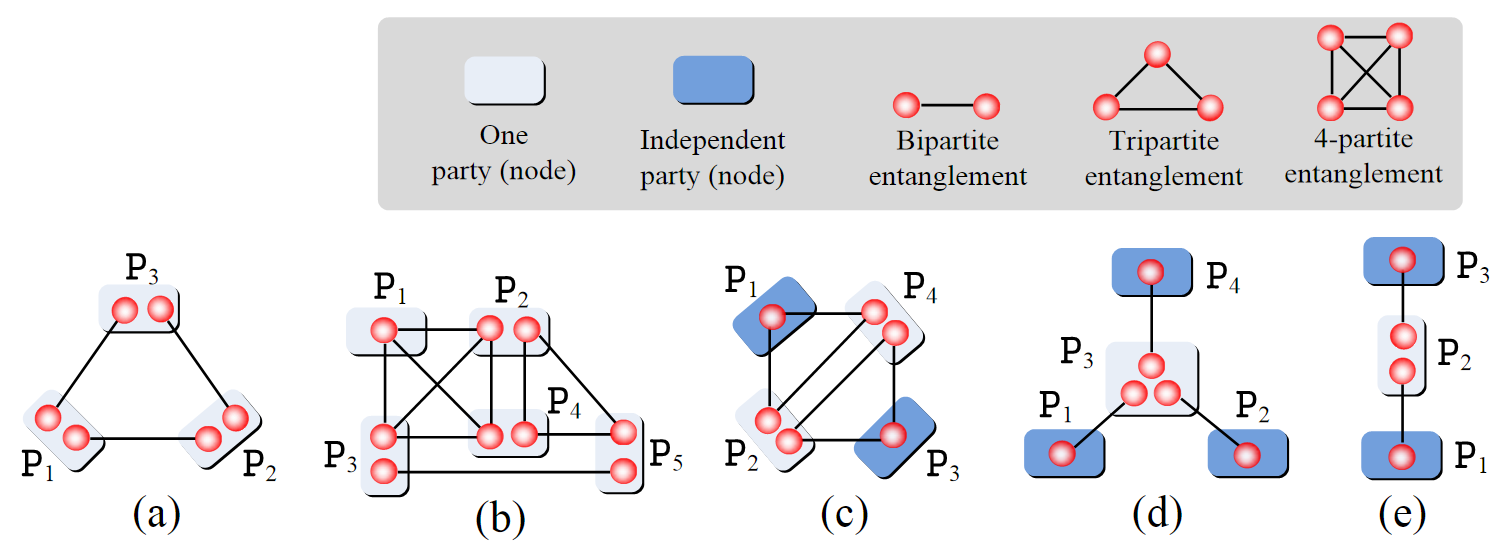}
\end{center}
\caption{(Color online) Schematic quantum network. (a) Triangle (cyclic) quantum network consisting of three observers $\texttt{P}_1, \texttt{P}_2, \texttt{P}_3$. $\texttt{P}_i$ and $\texttt{P}_{(i\mod 3)+1}$ share one bipartite entanglement, $i=1, 2, 3$. (b) Multiple cyclic quantum network consisting of five observers $\texttt{P}_1, \texttt{P}_2, \cdots, \texttt{P}_5$. The observers $\texttt{P}_1, \cdots, \texttt{P}_4$ share one 4-partite entanglement. $\texttt{P}_2, \texttt{P}_4, \texttt{P}_5$ share one tripartite entanglement. The observers $\texttt{P}_3, \texttt{P}_4, \texttt{P}_5$ share one tripartite entanglement. (c) Symmetric cyclic quantum network consisting of four observers $\texttt{P}_1, \cdots, \texttt{P}_4$. The observers $\texttt{P}_1, \texttt{P}_2, \texttt{P}_4$ share one tripartite entanglement. The observers $\texttt{P}_2, \texttt{P}_3, \texttt{P}_4$ share one tripartite entanglement. (d) Star-shaped quantum network consisting of four observers $\texttt{P}_1, \cdots, \texttt{P}_4$. Two observers $\texttt{P}_i$ and $\text{P}_3$ share one bipartite entanglement, $i=1, 2, 4$. (e) Chain-shaped quantum network consisting of three observers $\texttt{P}_1, \cdots, \texttt{P}_3$. Two observers $\texttt{P}_i$ and $\texttt{P}_2$ share one bipartite entanglement, $i=1, 3$. Networks in subfigures (a) and (b) have no independent observers while the networks in other subfigures have independent observers shown with blue boxes.
}
\end{figure}

\textbf{Corollary 4}. For a quantum network ${\cal N}_q$ consisting of $n+k+1$ observers $\texttt{P}_1, \cdots, \texttt{P}_{n+k}$, and Bob, assume that all observers $\texttt{P}_j$ with $j\in {\cal I}_i$ and Bob share one multipartite entanglement $\rho_{i}$ on the Hilbert space $(\otimes_{j\in {\cal I}_i} {\cal H}_{j})\otimes {\cal H}_{B_i}$, where ${\cal I}_i\subset\{1, \cdots, n+k\}$, $i=1, \cdots, s$. Then the joint system in the state $\rho_{1}\otimes\cdots \otimes \rho_{s}$ is $n+k+2$-partite nonlocal. Moreover, for the subnetwork consisting of $m$ observers $\texttt{P}_{j_1}, \cdots, \texttt{P}_{j_m}$, their reduced system is $m$-partite nonlocal after all the other observers performing a proper local POVM, where all observers of $\texttt{P}_{j_1}, \cdots, \texttt{P}_{j_m}$ are independent, i.e., they have not initially shared entangled states.

{\bf Proof}. The first part of Corollary 4, i.e., the $n+k+2$-partite locality is similar to that stated in Corollary 1, which can be proved by using the semiquantum nonlocal game for single entanglement \cite{SQ} and the procedures given in subsections A1 of Appendix A. Moreover, consider one subnetwork ${\cal N}_m$ consisting of independent observers $\texttt{P}_{j_1}, \cdots, \texttt{P}_{j_m}$, who have not initially shared entangled states. For each observer $\texttt{P}_{j_t}$, there is an integer set ${\cal I}_{l_{t}}$ satisfying $j_t\in {\cal I}_{l_{t}}$ and $j_r\not\in {\cal I}_{l_{t}}$ for $r\not=t$, i.e., two observers $\texttt{P}_{j_t}$ and Bob share an entanglement $\rho_{l_t}$. After a proper local POVM of all observers $\texttt{P}_{r}$ with $r\in {\cal I}_{l_{t}}$ and $r\not=j_t$, the observers $\texttt{P}_{j_t}$ and Bob can share a bipartite entanglement. Based on this fact, there is a standard star-shaped quantum subnetwork after a proper local POVM of all observers except for $\texttt{P}_{j_1}, \cdots, \texttt{P}_{j_m}$. The $m$-partite nonlocality of the subnetwork ${\cal N}_m$ is followed from Lemma 1. $\blacksquare$

In what follows, we complete the proof of Theorem 2.

{\bf Proof of Theorem 2}. Assume that a connected quantum network ${\cal N}_q$ consists of $n$ observers $\texttt{P}_1, \cdots, \texttt{P}_n$, who have shared entangled states $\rho_1, \cdots, \rho_m$, where each entanglement $\rho_i$ is shared by some observers $\texttt{P}_j$ with $j\in {\cal I}_i\subset \{1, 2, \cdots, n\}$, $i=1, \cdots, m$. From the definition of the connected quantum network, for each pair observers $\texttt{P}_s$ and $\texttt{P}_t$ there is one set of observers $\texttt{P}_{i_1}, \cdots, \texttt{P}_{i_k}$ such that any adjacent two observers of $\texttt{P}_s, \texttt{P}_{i_1}, \cdots, \texttt{P}_{i_k}, \texttt{P}_t$ share at least one entangled state. Based on this fact, we prove the result in three cases as follows.

Case 1. There is no independent observers in ${\cal N}_q$, i.e., any two observers have shared at least one entanglement, where some examples are shown in Figures S7(a) and S7(b). In this case, there are semiquantum nonlocal games $\mathbbm{G}^{(1)}_{sq}, \cdots, \mathbbm{G}^{(m)}_{sq}$ for entangled states $\rho_{1}, \cdots, \rho_m$, respectively \cite{SQ}. The $n$-partite nonlocality of ${\cal N}_q$ can be proved by iteratively using the procedure given in subsections A1 of Appendix A. The main steps are shown as follows:  we can firstly redefine a new semiquantum nonlocal game $\mathbbm{G}^{(1,2)}_{sq}$ to verify the joint system $\rho_1\otimes\rho_2$ from the semiquantum nonlocal games $\mathbbm{G}^{(1)}_{sq}$ and $\mathbbm{G}^{(2)}_{sq}$ using Corollary 1. And then, we redefine a new semiquantum nonlocal game $\mathbbm{G}^{(1,2,3)}_{sq}$ to verify the joint system $(\rho_1\otimes\rho_2)\otimes \rho_3$ from the semiquantum nonlocal games $\mathbbm{G}^{(1,2)}_{sq}$ and $\mathbbm{G}^{(3)}_{sq}$ using Corollary 1, where $\rho_1\otimes\rho_2$ is regarded as an entangled system. The $n$-partite nonlocality of $\rho_1\otimes\cdots\otimes\rho_m$ can be proved by repeating this procedure.

Case 2. There are some independent observers $\texttt{P}_i$ with $i\in {\cal I}\subset\{1, 2, \cdots, n\}$ in the network ${\cal N}_q$, i.e., any two observers $\texttt{P}_i$ with $i\in {\cal I}$ have not initially shared entanglement, where some examples are shown in Figures S7(c)-S7(e). Similar to Case 1, we can prove the $n$-partite nonlocality of ${\cal N}_q$ by repeating the procedure given in subsections A1 of Appendix A. The reason is from the definition of the connected quantum network.

\begin{figure}
\begin{center}
\includegraphics[width=.85\textwidth]{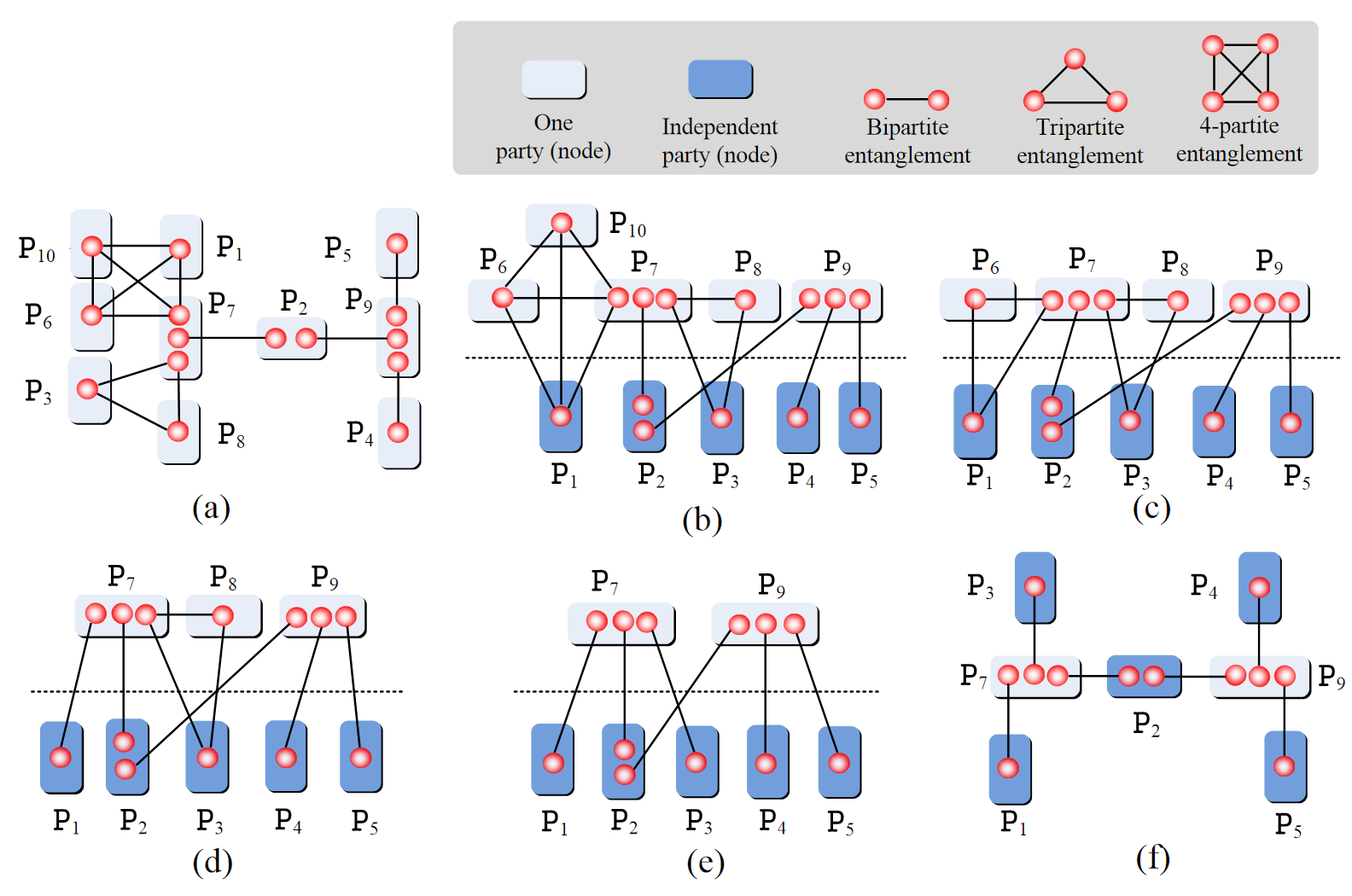}
\end{center}
\caption{(Color online) Schematic example of reducing a connected quantum network. (a) A connected quantum network consisting of ten observers $\texttt{P}_1, \cdots, \texttt{P}_{10}$. $\texttt{P}_i$ and $\texttt{P}_{9}$ share one bipartite entanglement, $i=2, 4, 5$. The observers $\texttt{P}_2$ and $\texttt{P}_{7}$ share one bipartite entanglement. The observers $\texttt{P}_3, \texttt{P}_7$ and $\texttt{P}_{8}$ share one tripartite entanglement. The observers $\texttt{P}_1, \texttt{P}_6, \texttt{P}_7$ and $\texttt{P}_{10}$ share one 4-partite entanglement. (b) An equivalent quantum network. Here, the observers $\texttt{P}_1, \cdots, \texttt{P}_5$ are independent observers, who have not initially shared entangled states. In the equivalent network, all the other observers $\texttt{P}_6, \cdots, \texttt{P}_{10}$ are placed above the dot line. (c) Reduced quantum network of the network shown in Figure S8(b). To obtain this quantum network, we consider the quantum subnetwork consisting of four observers $\texttt{P}_1, \texttt{P}_2, \texttt{P}_{6}$, $\texttt{P}_7$ ($j_1=6$, $j_2=7$ for the proof in Case 3 shown in Figure S8(b). Since $j_1=6\not=j_2=7$, there is a chain-shaped quantum subnetwork ${\cal N}_{1,2}$ consisting of two observers $\texttt{P}_6, \texttt{P}_{7}$ as shown in Figure S8(b). Note that there is another observer $\texttt{P}_{10}$ who shares the same system with the observers $\texttt{P}_6$ and $\texttt{P}_{7}$. So, we obtain a new quantum network after the observer $\texttt{P}_{10}$ performing a local POVM to disentangle his shared 4-partite entanglement into a tripartite entanglement. (d) Reduced quantum network of the network shown in Figure S8(c). The new quantum network is obtained after the observer $\texttt{P}_{6}$ performing a local POVM to disentangle his shared tripartite entanglement into a bipartite entanglement. We cannot require the observer $\texttt{P}_{7}$ to disentangle the tripartite entanglement shared with three observers $\texttt{P}_1, \texttt{P}_6$ and $\texttt{P}_7$. The reason is that the reduced quantum network should be connected. (e) Reduced quantum network of the network shown in Figure S8(d). Now, we consider a new quantum subnetwork ${\cal N}_{2,3}$ of the network shown in Figure S8(d), which consists of four observers $\texttt{P}_2, \texttt{P}_3, \texttt{P}_7, \texttt{P}_8$. The reduced quantum network is obtained after the observer $\texttt{P}_{8}$ performing a local POVM to disentangle the tripartite entanglement shared with three observers $\texttt{P}_7, \texttt{P}_8$ and $\texttt{P}_3$ into a bipartite entanglement, where $\texttt{P}_8\not \in{\cal N}_{2,3}\cup\{\texttt{P}_4\}$. (f) Equivalent quantum network of the network shown in Figure S8(e). It is a hybrid quantum network of a chain-shaped quantum subnetwork consisting of three observers $\texttt{P}_7, \texttt{P}_2$, and $\texttt{P}_9$, and two star-shaped quantum subnetworks consisting of four observers $\texttt{P}_1, \texttt{P}_2$, $\texttt{P}_3$, and $\texttt{P}_7$, or $\texttt{P}_2, \texttt{P}_4$, $\texttt{P}_5$, and $\texttt{P}_9$. All independent observers are shown with blue boxes. }
\end{figure}

Case 3. There are some independent observers $\texttt{P}_i$ with $i\in {\cal I}\subset\{1, 2, \cdots, n\}$ in the quantum network ${\cal N}_q$. Different from Case 2, we can further prove the activated nonlocality of any subnetwork consisting of all observers $\texttt{P}_i$ with $i\in {\cal I}'\subseteq {\cal I}$, where an example is shown in Figure S8(a). In detail, there is a subnetwork ${\cal N}_q'$ consisting of all observes $\texttt{P}_i$ with $i\in {\cal I}'$, which is obtained by measuring the joint system $\rho_1\otimes \cdots\otimes \rho_m$ with a proper local POVM of all the other observers. Moreover, the reduced subnetwork consists of several chain-shaped and star-shaped quantum subnetworks. The proof of this fact is iteratively followed by using an equivalent schematic network, where an example is shown in Figure S8(b). For convenience, assume that all observers $\texttt{P}_1, \cdots, \texttt{P}_k$ are independent, where ${\cal I}'=\{1, \cdots, k\}$ with $k<n$. Moreover, assume that each pair observers $\texttt{P}_i$ and $\texttt{P}_{j_i}$ share an entanglement $\rho_i$, where some integers $j_s, j_t$ may satisfy $j_s=j_t$ for $s\not=t$, see the star-shaped quantum network shown in Figure S6. Note that the assumption of shared entangled states is reasonable from the independence of all observers $\texttt{P}_1, \cdots, \texttt{P}_k$. Now, we can reduce this subnetwork ${\cal N}'$ as follows.

Take the subnetwork consisting of the observers $\texttt{P}_1, \texttt{P}_2, \texttt{P}_{j_1}, \texttt{P}_{j_2}$ as an example.
\begin{itemize}
\item[S1.] When $j_1=j_2$, the joint system $\rho_1\otimes \rho_2$ consists of a generalized $\Lambda$-shaped quantum network. In this subcase, after all observers who share the systems $\rho_1\otimes \rho_2$ except for $\texttt{P}_1, \texttt{P}_2, \texttt{P}_{j_1}$ performing a proper local POVM on the joint system $\rho_1\otimes \rho_2$, the reduced system shared by the observer $\texttt{P}_1, \texttt{P}_2, \texttt{P}_{j_1}$ is a standard $\Lambda$-shaped quantum network consisting of entangled states. Note that all these local POVMs do not change the joint system $\rho_3\otimes \cdots \otimes \rho_m$ (see the subnetwork consisting of three observers $\texttt{P}_4$, $\texttt{P}_5$, and $\texttt{P}_{j_4}$ shown in Figures S8(c) and S8(d) as an example).
\item[S2.] When $j_1\not=j_2$, there is a generalized chain-shaped quantum subnetwork ${\cal N}_{1,2}$ consisting of the observers $\texttt{P}_{j_1}, \texttt{P}_{t_1}, \cdots, \texttt{P}_{t_{s-1}}, \texttt{P}_{j_2}$ from the assumption of the connected quantum network, where any adjacent two observers in ${\cal N}_{1,2}$ share at least one entangled state. Assume that ${\cal N}_{1,2}$ consists of a joint system $\rho_{t_1}\otimes \cdots \otimes \rho_{t_s}$, where the observers $\texttt{P}_{j_1}$ and $\texttt{P}_{t_1}$ share the entanglement $\rho_{t_1}$, $\texttt{P}_{t_l}$ and $\texttt{P}_{t_{l+1}}$ share the entanglement $\rho_{t_{l+1}}$, and $\texttt{P}_{t_s}$ and $\texttt{P}_{j_2}$ share the entanglement $\rho_{t_{s}}$, and $t_l\not=1,2$ for all $l$s. For each entanglement $\rho_{t_l}$ that is also shared by some observer $\texttt{P}_i$ with $i>k$, the observer $\texttt{P}_{t_l}$ or $\texttt{P}_i$ can perform a disentangling operation such that the reduced joint system is entangled, where the reduced quantum network should be connected after the disentangling operations. Moreover, for each entanglement $\rho_{t_l}$ that is shared by some observer $\texttt{P}_i$ with $i\leq k$, the observer $\texttt{P}_{t_l}$ can perform a disentangling operation such that the reduced joint system is entangled, where the reduced quantum network should be connected. After all observers who do not belong to ${\cal N}_{1,2}\cup\{\texttt{P}_3, \cdots, \texttt{P}_k\}$ performing a local POVM on the joint system $\rho_{t_1}\otimes \cdots \otimes \rho_{t_s}$, the reduced subnetwork ${\cal N}_{1,2}$ is a standard chain-shaped quantum network shown in Figure S5. If $t_l\not\in \{3, \cdots, m\}$ for all $l$s, all these local POVMs do not change the joint system $\rho_3\otimes \cdots \otimes \rho_m$. Otherwise, each entanglement $\rho_{j}$ with $t_l=j$ for some $l$ is changed into a new entanglement $\hat{\rho}_{j}$, which is shared by two observers $\texttt{P}_j$ and $\texttt{P}_{t_{l-1}}$, where $t_{0}=j_1$ and $t_s=j_2$. One example is shown in Figures S8.
\end{itemize}

Now, similar reducing procedure can be performed for the subnetwork consisting of the subnetwork ${\cal N}_{1,2}$ and the parties $\texttt{P}_3, \texttt{P}_{j_3}$. By iteratively repeating these reducing procedures of subcases 1 and 2 for all observers $\texttt{P}_1, \texttt{P}_1, \texttt{P}_{j_1}, \texttt{P}_{j_2}$, we can obtain a generalized subnetwork ${\cal N}'$ consisting of star-shaped quantum subnetworks and chain-shaped quantum subnetworks, seeing one example shown in Figures S8(e) and S8(f), where all entangled states thate are not involved in these reducing procedures can be measured with any local POVMs.

In what follows, we show the $k$-partite nonlocality of each generalized hybrid quantum network (one example is shown in Figure S8(f)). It can be completed by combining Corollary 2 and Lemma 1. Specially, we can firstly prove the multipartite nonlocality for all star-shaped quantum subnetworks from Lemma 1. And then we can complete the proof for the hybrid quantum network shown in Figure S8(f) from Corollary 2, where the procedure given in subsections A1 of Appendix A can be iteratively used. Consequently, we have proved Theorem 2. $\blacksquare$

\section*{Appendix C: Proof of Theorem 3}

In this section, we prove the nonlocality of a hybrid network, see Figure 3. In detail, we consider an arbitrary quantum network ${\cal N}_q$ consisting of $n$ observers $\texttt{P}_1, \cdots, \texttt{P}_n$ with $n\geq 3$. Assume that the quantum resources of ${\cal N}$ consist of $\rho_1\otimes\cdots \otimes\rho_m$. When all $\rho_i$s are bipartite or multipartite entangled states that are shared by two or multiple observers. ${\cal N}_q$ is reduced to the network shown in Theorem 2 if it is connected. Now, we consider the following case. Note that ${\cal N}_q$ can be regarded as a classical network that is local when all $\rho_i$s are fully separable. So, in what follows, assume that there are some $\rho_i$s that are entangled. We can prove that ${\cal N}_q$ is $n$-partite nonlocal even if ${\cal N}_q$ is disconnected. Formally, we prove the result shown in Theorem 3. The proof is followed from Theorem 2. In fact, assume that ${\cal N}_q$ consists of $k$ connected quantum subnetworks $\tilde{\cal N}_1, \cdots, \tilde{\cal N}_k$, and classical network $\hat{\cal N}$, i.e.,
\begin{align*}
{\cal N}=(\cup_{i=1}^k\tilde{\cal N}_i)\cup\hat{\cal N},
\tag{C1}
\end{align*}
where $\tilde{\cal N}_i$ and $\tilde{\cal N}_j$ have no shared observer for any $i\not=j$, and $\tilde{\cal N}_i$ and $\hat{\cal N}$ have some shared observers. One example is shown in Figure 3(a). Here, $\tilde{\cal N}_i$ consists of all entangled states for all $i$s while $\hat{\cal N}$ consists of all fully separable states.

For each connected quantum subnetwork $\tilde{\cal N}_i$ consisting of $n_i$ observers with a quantum state $\varrho_i$, from Theorem 2 there is a semiquantum nonlocal game $\mathbbm{G}^{(i)}_{sq}$ (or generalized Bell-type inequalities) such that the average expect $\wp_i$ satisfies
\begin{align*}
\wp^{(i)}_q(\varrho_i)>c_i,
\tag{C2}\\
\sup_{\hat{Q}_1, \cdots \hat{Q}_{n_i}}\wp^{(i)}_c(\varrho^c_i)\leq c_i,
\tag{C3}
\end{align*}
where $\varrho^c_i$ denotes the classical resource of the subnetwork $\tilde{\cal N}_i$ (or the fully separable states), $c_i\geq0$ denotes the available classical upper bound of the average gain $\wp^{(i)}_c(\varrho^c_i)$ which depends on the multipartite classical correlations from $\tilde{\cal N}_i$ in terms of all local POVMs $\hat{Q}_1, \cdots, \hat{Q}_{n_i}$ for all observers in $\tilde{\cal N}_i$. The inequality (C2) means that there are some local POVMs $Q_1, \cdots Q_{n_i}$  for all observers in $\tilde{\cal N}_i$ such that $\wp^{(i)}_q(\varrho_i)>c_i$ for the quantum state $\varrho_i$.

For the classical subnetwork $\hat{\cal N}$ consisting of $n'$ observers with a quantum state $\hat{\varrho}$, for each semiquantum nonlocal game $\hat{\mathbbm{G}}_{sq}$ there are some local POVMs $\hat{Q}'_{1}, \cdots, \hat{Q}'_{n'}$ such that
\begin{align*}
\hat{\wp}_c(\hat{\varrho})=\hat{c}
\tag{C4}
\end{align*}
where $\hat{c}=\sup_{\hat{Q}_1, \cdots \hat{Q}_{n'}}\hat{\wp}_c(\hat{\varrho})\geq 0$. From forward computations using the similar reconstruction procedure shown in subsection A1 of Appendix A, it is easy to prove that there is a global semiquantum nonlocal game $\mathbbm{G}_{sq}$ such that
\begin{align*}
\wp_q(\rho_1\otimes \cdots \otimes \rho_m)>c,
\tag{C5}
\\
\sup_{\hat{Q}_1, \cdots \hat{Q}_{n}}\wp_c(\varrho_1\otimes \cdots \otimes \varrho_m)\leq c
\tag{C6}
\end{align*}
where $\wp$ is the average gain \cite{SQ} depending on the joint conditional probabilities of the measurement outcomes of all observers, $\rho_i$s are multipartite states satisfying that there is at least one entangled state $\rho_i$, all $\varrho_j$s are fully separable states. The inequality (C5) means that there are some local POVMs ${Q}_1, \cdots Q_{n}$ of all observers such that the average gain $\wp$ from quantum multipartite correlations of a hybrid quantum network is no less than a constant $c$. The inequality (C6) means that for all local POVMs $\hat{Q}_1, \cdots \hat{Q}_{n}$ of all observers the maximal average gain $\wp$ from multipartite correlations of a classic network consisting of all fully separable states is no more than the constant $c$. Here, the inequalities (C5) and (C6) can be regarded as generalized Bell-type inequalities. It follows that any hybrid quantum network consisting of at least one entangled state has multipartite nonlocality. We have proved Theorem 3. $\blacksquare$

\section*{Appendix D: Verifying entanglement swapping without LOCC}

In this section, we present some discussions related to verifying entanglement swapping with one copy of joint system. In detail, consider a $\Lambda$-shaped quantum network consisting of two bipartite entangled states $\rho_{AB}\otimes \rho_{CD}$ shared by three parties, where Alice and Bob share the entanglement $\rho_{AB}$ while Bob and Charlie share the entanglement $\rho_{CD}$. Theorem 1 shows that there is a generalized Bell-type inequality for verifying the bipartite activated nonlocality of Alice and Charlie after a local POVM of Bob from $\rho_{AB}^{\otimes n}\otimes \rho_{CD}^{\otimes n}$ with large $n$. In this section, we present some result for $\rho_{AB}\otimes \rho_{CD}$.

To explain the main idea, assume that $\rho_{AB}$ and $\rho_{CD}$ has the following forms:
\begin{align*}
\rho_{AB}=p_1|\Phi_1\rangle_{AB}\langle \Phi_1|+p_2|\Phi_2\rangle_{AB}\langle \Phi_2|,
\tag{D1}
\\
\rho_{CD}=q_1|\Psi_1\rangle_{CD}\langle \Psi_1|+q_2|\Psi_2\rangle_{CD}\langle \Psi_2|,
\tag{D2}
\end{align*}
where $|\Phi_i\rangle_{AB}, |\Psi_i\rangle_{CD}$ are assumed to be bipartite pure states.

Case 1. $|\Phi_i\rangle_{AB}, |\Psi_i\rangle_{CD}$ are all bipartite entangled states with $i=1,2$. In this case, assume that $|\Phi_1\rangle_{AB}=\sum_{i}a_i|ii\rangle_{AB}$ and $|\Psi_1\rangle_{AB}=\sum_{j}b_j|jj\rangle_{CD}$ which are formal bipartite entangled states, and $|\Phi_2\rangle_{AB}=(U_A\otimes U_B)\sum_{i}a'_i|ii\rangle_{AB}$ and $|\Psi_2\rangle_{CD}=(U_C\otimes U_D)\sum_{j}b'_j|jj\rangle_{CD}$ which are formal bipartite entangled states under the local unitary transformations $U_A\otimes U_B, U_C\otimes U_D$, respectively. Here, $U_A, U_B, U_C, U_D$ are local unitary operations performed on the systems $A, B, C, D$, respectively. Note that for each joint system: $\rho_{ij}=|\Phi_i\rangle_{AB}|\Psi_j\rangle_{CD}\langle \Psi_{j}|\langle \Phi_i|$, CHSH inequality is useful for verifying the bipartite activated nonlocality of Alice and Charlie after a Bell measurement of Bob \cite{CHSH,Gisin1}, where we can take use of the equality $\textrm{Tr}[M_{BC}\rho_{ij}]=\textrm{Tr}[M_{BC}(W_{A}\otimes W_{BC}\otimes W_{D}) \rho_{ij}(W_{A}\otimes W_{BC}\otimes W_{D})^\dag]=\textrm{Tr}[M_{BC} W_{BC} \rho_{ij} W_{BC}^\dag]$, $W_{A}$, $W_{BC}$, and $W_{D}$ are unitary operations on the systems $A, BC, D$, respectively, and $M_{BC}$ is a local POVM on the systems $BC$. Combining the equality $\rho_{AB}\otimes \rho_{CD}=\sum_{i,j}p_iq_j\rho_{ij}$, it follows that the bipartite activated nonlocality of Alice and Charlie for the joint system $\rho_{AB}\otimes \rho_{CD}$ may be verified by using CHSH inequality \cite{CHSH,Gisin1} after a Bell measurement of Bob. Generally, we cannot obtain the deterministic detecting of the bipartite activated nonlocality because the local unitary operations $U_A, U_B, U_C, U_D$ can affect the upper bounds $c_{ij}$s, see Example 1. Similar result holds for general bipartite entangled states $\rho_{AB}\otimes \rho_{CD}$, where $\rho_{AB}=\sum_ip_i|\Phi_i\rangle_{AB}\langle \Phi_i|$, $\rho_{CD}=\sum_jq_j|\Psi_j\rangle_{CD}\langle\Psi_j|$, and  $|\Phi_i\rangle_{AB}, |\Psi_i\rangle_{CD}$ are bipartite entangled states for all $i,j$s.

Case 2. $|\Phi_1\rangle_{AB}, |\Psi_i\rangle_{CD}$ are all bipartite entangled states with $i=1, 2$. Here, $|\Phi_2\rangle_{AB}$ is separable state. In this case, assume that $|\Phi_1\rangle_{AB}=\sum_{i}a_i|ii\rangle_{AB}$ and $|\Psi_1\rangle_{AB}=\sum_{j}b_j|jj\rangle_{CD}$ which are formal bipartite entangled states, and $|\Psi_2\rangle_{CD}=(U_C\otimes U_D)\sum_{j}b'_j|jj\rangle_{CD}$ which is formal bipartite entanglement under the local unitary transformations $U_C\otimes U_D$, respectively. Here, $U_C, U_D$ are local unitary operations performed on the system $C, D$, respectively. Define
\begin{align*}
\rho_{ij}=|\Phi_i\rangle_{AB}|\Psi_j\rangle_{CD}\langle \Psi_{j}|\langle \Phi_i|, i,j=1,2.
\tag{D4}
\end{align*}
Assume that a Bell-type inequality which is linearly depending on bipartite correlations $P(\textbf{x}|\textbf{a})$ is given as follows
\begin{align*}
\wp(P_q(\textbf{x}|\textbf{a}))>c,
\tag{D5}
\\
\wp(P_c(\textbf{x}|\textbf{a}))\leq c,
\tag{D6}
\end{align*}
where $P_q(\textbf{x}|\textbf{a})$ and $P_c(\textbf{x}|\textbf{a})$ are the joint conditional probability distributions from the local measurements of an entangled system $\rho_{AD}$, and fully separable systems or classical states $\varrho_{AD}$, respectively. $c$ denotes the maximal classical bound over any possible conditional probabilities $P_c(\textbf{x}|\textbf{a})$s. Assume that there are local POVMs of three observers, i.e., ${\mathscr{X}}_1$-POVM $\{P^{x_1}\}\in {\cal M}_{A; {\mathscr{X}}_1}$ for Alice, ${\mathscr{X}}_2$-POVM $\{Q^{x_2}\}\in {\cal M}_{BC; {\mathscr{X}}_2}$ for Bob, and ${\mathscr{X}}_3$-POVM $\{R^{x_3}\}\in {\cal M}_{D; {\mathscr{X}}_3}$ for Charlie. For each joint system $\rho_{ij}$, define
\begin{align*}
\wp(P_{ij}(\textbf{x}|\textbf{a}))=c_{ij}
\tag{D7}
\end{align*}
where $P_{ij}(\textbf{x}|\textbf{a})$ denotes the joint conditional probability distributions from the local POVMs of Alice and Charlie after a local POVM by Bob on his systems $BC$ of $\rho_{ij}$. It follows that
\begin{align*}
\wp(P_q(\textbf{x}|\textbf{a}))=&\wp(\sum_{ij}p_iq_jP_{ij}(\textbf{x}|\textbf{a}))
\\
=&\sum_{ij}p_iq_jc_{ij},
\tag{D8}
\end{align*}
where the equation (D8) is from the linearity of $\wp$. So, in order to verify the bipartite activated nonlocality, we have
\begin{align*}
\sum_{ij}p_iq_jc_{ij}>c
\tag{D9}
\end{align*}
for some POVMs of all observers, where $c$ is given in the equation (D6). Here, from assumptions $\rho_{11}$ and $\rho_{12}$ are useful joint systems. $\rho_{21}$ and $\rho_{22}$ are useless even if they are not classical system, where $|\Psi_i\rangle_{CD}$ are entangled. So, we can obtain the following equivalent inequality of (D9) as
\begin{align*}
q_1c_{11}+q_2c_{12}>\frac{1}{p_1}(p_2(q_1c_{11}+q_2c_{12})-c)
\tag{D10}
\end{align*}
which should be satisfied by a joint system $\rho_{AB}\otimes \rho_{CD}$ if it is bipartite activated nonlocal in terms of a Bell inequality shown in the equation (D6). Similar result holds for general bipartite entangled states $\rho_{AB}\otimes \rho_{CD}$ consisting of multiple bipartite entangled pure states.

In what follows, we present some examples.

\textbf{Example 1}. Assume that a $\Lambda$-shaped quantum network consists of two bipartite entangled states $\rho_{AB}$ and $\rho_{CD}$, where
\begin{align*}
\rho_{AB}=p_1|\Phi_1\rangle\langle \Phi_1|+p_2|\Phi_2\rangle\langle \Phi_2|,
\tag{D11}
\\
\rho_{CD}=q_1|\Psi_1\rangle\langle \Psi_1|+q_2|\Psi_2\rangle\langle \Psi_2|,
\tag{D12}
\end{align*}
$|\Phi_i\rangle$ and $|\Psi_i\rangle$ are bipartite entangled pure states which are defined by
\begin{align*}
|\Phi_1\rangle=&a_1|00\rangle+b_1|11\rangle,
\tag{D13}\\
|\Psi_1\rangle=&a_2|00\rangle+b_2|11\rangle,
\tag{D14}\\
|\Phi_2\rangle=&c_1|01\rangle+c_1|10\rangle,
\tag{D15}\\
|\Psi_2\rangle=&d_2|01\rangle+d_2|10\rangle,
\tag{D16}
\end{align*}
$p_1+p_2=q_1+q_2=a_i^2+b_i^2=c_i^2+d_i^2=1$, $0\leq p_i,q_i\leq 1$, $0< a_i,b_i,c_i,d_i<1$, $i=1,2$. Now, consider bipartite CHSH inequality \cite{CHSH} as
\begin{align*}
\wp(P(\textbf{x}|\textbf{a})):=&\langle {M}^{1}_0\otimes {M}^2_0\rangle
+\langle{M}^{1}_0\otimes {M}^2_1\rangle
+\langle{M}^{1}_1\otimes{M}^2_0\rangle
-\langle{M}^{1}_1\otimes{M}^2_1\rangle
\\
\leq & 2,
\tag{B17}
\end{align*}
where ${M}^{1}_i, {M}^{2}_j$ denote dichotomic quantum observables with $\pm 1$ outputs. Now, define dichotomic quantum observables $M^1_i=(1-i)\sigma_z+i\sigma_x$, $i=0, 1$, and $M^2_i=\cos\theta\sigma_z+(-1)^i\sin\theta \sigma_x$, where $\sigma_z$ and $\sigma_x$ are Pauli matrices. Assume that Bob performs Bell measurement on the system $BC$ under the basis $\{\frac{1}{\sqrt{2}}(|00\rangle\pm |11\rangle), \frac{1}{\sqrt{2}}(|01\rangle\pm |10\rangle)\}$. In the following, we only compute one local measurement of Bob with POVM $M_{BC}=\frac{1}{2}\sum_{i,j=0,1}|ii\rangle\langle jj|$.
It is easy to follow that
\begin{align*}
c_{11}=&\textrm{Tr}[({M}^{1}_0\otimes {M}^2_0+{M}^{1}_0\otimes {M}^2_1
+{M}^{1}_1\otimes{M}^2_0-{M}^{1}_1\otimes{M}^2_1)\hat{\rho}_{11}]
\\
=&2\cos(\theta)+4\frac{a_1b_1a_2b_2}{a_1^2a_2^2+b_1^2b_2^2}\sin(\theta),
\tag{D18}\\
c_{12}=&\textrm{Tr}[({M}^{1}_0\otimes {M}^2_0+{M}^{1}_0\otimes {M}^2_1
+{M}^{1}_1\otimes{M}^2_0-{M}^{1}_1\otimes{M}^2_1)\hat{\rho}_{12}]
\\
=&-2\cos(\theta)+4\frac{a_1b_1c_2d_2}{a_1^2c_2^2+b_1^2d_2^2}\sin(\theta),
\tag{D19}\\
c_{21}=&\textrm{Tr}[({M}^{1}_0\otimes {M}^2_0+{M}^{1}_0\otimes {M}^2_1
+{M}^{1}_1\otimes{M}^2_0-{M}^{1}_1\otimes{M}^2_1)\hat{\rho}_{21}]
\\
=&
-2\cos(\theta)+4\frac{c_1d_1a_2b_2}{c_1^2a_2^2+d_1^2b_2^2}\sin(\theta),
\tag{D20}\\
c_{14}=&\textrm{Tr}[({M}^{1}_0\otimes {M}^2_0+{M}^{1}_0\otimes {M}^2_1
+{M}^{1}_1\otimes{M}^2_0-{M}^{1}_1\otimes{M}^2_1)\hat{\rho}_{22}]
\\
=&2\cos(\theta)+4\frac{c_1d_1c_2d_2}{c_1^2c_2^2+d_1^2d_2^2}\sin(\theta),
\tag{D21}
\end{align*}
where $\hat{\rho}_{ij}=\textrm{Tr}_{BC}[M_{BC}|\Phi_i\rangle_{AB}|\Psi_j\rangle_{CD}
\langle\Psi_j|\langle\Phi_i|]/\textrm{Tr}[M_{BC}|\Phi_i\rangle_{AB}|\Psi_j\rangle_{CD}
\langle \Psi_j|\langle\Phi_i|]$, $\textrm{Tr}_{BC}$ denotes the partial trace operations performed on the systems $BC$. From the equations (D18)-(D21), it follows from the equation (D9) that
\begin{align*}
\sum_{i,j=1,2}p_iq_jc_{ij}=&2\alpha\cos(\theta)+2\beta\sin(\theta)
\\
=&2\sqrt{\alpha^2+\beta^2}
\\
>&2
\tag{D22}
\end{align*}
for some $a_i,b_i,c_i,d_i,p_i,q_i$s, where $\cos(\theta):=\alpha/\sqrt{\alpha^2+\beta^2}$, $\alpha=(p_1-p_2)(q_1-q_2)$, and $\beta=p_1q_1\frac{2a_1b_1a_2b_2}{a_1^2a_2^2+b_1^2b_2^2}
+p_1q_2\frac{2a_1b_1c_2d_2}{a_1^2c_2^2+b_1^2d_2^2}
+p_2q_1\frac{2c_1d_1a_2b_2}{c_1^2a_2^2+d_1^2b_2^2}
+p_2q_2\frac{c_1d_1c_2d_2}{c_1^2c_2^2+d_1^2d_2^2}\leq 1$ from the inequalities $2a_1b_1a_2b_2\leq a_1^2a_2^2+b_1^2b_2^2$, $2a_1b_1c_2d_2\leq a_1^2c_2^2+b_1^2d_2^2$, $2c_1d_1a_2b_2\leq c_1^2a_2^2+d_1^2b_2^2$ and $2c_1d_1c_2d_2\leq c_1^2c_2^2+d_1^2d_2^2$. Note that the inequality (D22) holds for special quantum states such as $a_1=b_1,c_1=d_1$ and $a_2d_2=b_2c_2$.

\textbf{Example 2}. Assume that a $\Lambda$-shaped quantum network consists of two bipartite Werner states as:
\begin{align*}
\rho_{AB}=\frac{1-p}{4}\mathbb{I}_4+p|\Phi\rangle_{AB}\langle \Phi|,
\tag{D23}
\\
\rho_{CD}=\frac{1-q}{4}\mathbb{I}_4+q|\Psi\rangle_{CD}\langle \Psi|,
\tag{D24}
\end{align*}
where $|\Phi\rangle_{AB}=a_1|00\rangle+b_1|11\rangle$, $|\Psi\rangle_{CD}=a_2|00\rangle+b_2|11\rangle$, $\mathbb{I}_4$ as the identity matrix of rank $4$ can be viewed as a mixed fully separable state, and $0\leq p, q\leq 1$. Define $\rho_{11}=\frac{1}{16}\mathbb{I}_4\otimes \mathbb{I}_4$, $\rho_{12}=\frac{1}{4}\mathbb{I}_4\otimes |\Psi\rangle_{CD}\langle \Psi|$, $\rho_{21}=\frac{1}{4}|\Phi\rangle_{AB}\langle \Phi|\otimes \mathbb{I}_4$, and $\rho_{22}=|\Phi\rangle_{AB}\langle \Phi|\otimes |\Psi\rangle_{CD}\langle \Psi|$. Let $\hat{\rho}_{ij}=\textrm{Tr}_{BC}[M_{BC}\rho_{ij}]/\textrm{Tr}[M_{BC}\rho_{ij}]$
for $i,j=1,2$, where $M_{BC}$ denotes the local POVM of Bob.

Similar to Example 1 with the same quantum observables for three observers, we can easily obtain
\begin{align*}
c_{11}=&\textrm{Tr}[({M}^{1}_0\otimes {M}^2_0+{M}^{1}_0\otimes {M}^2_1
+{M}^{1}_1\otimes{M}^2_0-{M}^{1}_1\otimes{M}^2_1)\hat{\rho}_{11}]
\\
=&0,
\tag{D25}\\
c_{12}=&\textrm{Tr}[({M}^{1}_0\otimes {M}^2_0+{M}^{1}_0\otimes {M}^2_1
+{M}^{1}_1\otimes{M}^2_0-{M}^{1}_1\otimes{M}^2_1)\hat{\rho}_{12}]
\\
=&0,
\tag{D26}\\
c_{21}=&\textrm{Tr}[({M}^{1}_0\otimes {M}^2_0+{M}^{1}_0\otimes {M}^2_1
+{M}^{1}_1\otimes{M}^2_0-{M}^{1}_1\otimes{M}^2_1)\hat{\rho}_{21}]
\\
=&0,
\tag{D27}\\
c_{14}=&\textrm{Tr}[({M}^{1}_0\otimes {M}^2_0+{M}^{1}_0\otimes {M}^2_1
+{M}^{1}_1\otimes{M}^2_0-{M}^{1}_1\otimes{M}^2_1)\hat{\rho}_{22}]
\\
=&2\cos(\theta)+4\frac{a_1b_1a_2b_2}{a_1^2a_2^2+b_1^2b_2^2}\sin(\theta),
\tag{D28}
\end{align*}
From the equations (D25)-(D28), it follows from the equation (D9) or (D10) that
\begin{align*}
\sum_{i,j=1,2}p_iq_jc_{ij}=&2pq(\cos(\theta)+\beta\sin(\theta))
\\
=&2pq\sqrt{1+\beta^2}
\\
>&2
\tag{D29}
\end{align*}
for some $a_i,b_i,c_i,d_i,p,q$s, where $\cos(\theta):=1/\sqrt{1+\beta^2}$, and $\beta=2a_1b_1a_2b_2/(a_1^2a_2^2+b_1^2b_2^2)\leq 1$. For example, the inequality (D29) holds for the following condition
\begin{align*}
pq>\frac{1}{\sqrt{1+\beta^2}}.
\tag{D30}
\end{align*}
Specially, when $\beta=1$, i.e., $2a_1b_1a_2b_2=a_1^2a_2^2+b_1^2b_2^2$ including $a_i=b_i=1/\sqrt{2}$ (Werner state of EPR state), it follows from the inequality (D30) that $p>\frac{1}{\sqrt{2}}$ and $q=1$, or $q>\frac{1}{\sqrt{2}}$ and $p=1$.


\begin{thebibliography}{99}
\bibitem{von} J. von Neumann, \textit{Mathematische Grundlagen der Quantenmechanik}, Springer, Berlin, 1932.

\bibitem{EPR1}A. Einstein, B. Podolsky, and N. Rosen, Can quantum-mechanical description of physical reality be considered complete? \textit{Phys. Rev.} \textbf{47}, 777 (1935).

\bibitem{Bell}J. S. Bell, On the Einstein-Podolsky-Rosen paradox, \textit{Phys}. \textbf{1}, 195 (1964).

\bibitem{CHSH}J. F. Clauser,  M. A. Horne,  A.  Shimony, and R. A. Holt, Proposed experiment to test local hidden-variable theories, \textit{Phys. Rev. Lett.} \textbf{23}, 880 (1969).

\bibitem{Gisin1}N. Gisin, Bell's inequality holds for all non-product states, \textit{Phys. Lett. A} \textbf{154}, 201 (1991).

\bibitem{GP}N. Gisin and A. Peres, Maximal violation of Bell's inequality for arbitrarily large spin, \textit{Phys. Lett. A} \textbf{162}, 15 (1992).

\bibitem{PR}S. Popescu and  D. Rohrlich, Generic quantum nonlocality, \textit{Phys. Lett. A} \textbf{166}, 293 (1992).

\bibitem{Pop}S. Popescu, Bell's inequalities versus Teleportation. What is nonlocality? \textit{Phys. Rev. Lett.} \textbf{72}, 797 (1994).

\bibitem{LF}M. Li and S.-M. Fei, Gisin's theorem for arbitrary dimensional multipartite states, \textit{Phys. Rev. Lett}. \textbf{104}, 240502 (2010).

\bibitem{Hardy}L. Hardy, Quantum mechanics, local realistic theories, and Lorentz-invariant realistic theories, \textit{Phys. Rev. Lett.} \textbf{68}, 2981 (1992).

\bibitem{YCZ}S. Yu, Q. Chen, C. Zhang, C. H. Lai, and C. H. Oh, All entangled pure states violate a single Bell's inequality, \textit{Phys. Rev. Lett}. \textbf{109}, 120402 (2012).

\bibitem{SQ}F. Buscemi, All entangled quantum states are nonlocal, \textit{Phys. Rev. Lett.} \textbf{108}, 200401 (2012).

\bibitem{GH}N. Gisin and B. Huttner, Quantum cloning, eavesdropping and Bell's inequality, \textit{Phys. Lett. A} \textbf{228}, 463 (1997).

\bibitem{Masa}L. Masanes, All entangled states are useful for information processing, \textit{Phys. Rev. Lett.} \textbf{96}, 150501 (2006).

\bibitem{PAMB} S. Pironio, A. Ac\'{\i}n, S. Massar, A. Boyer de la Giroday, D. N. Matsukevich, P. Maunz, S. Olmschenk, D. Hayes, L. Luo, T. A. Manning, and C. Monroe, Random numbers certified by Bell's theorem, \textit{Nature} (London) {\bf 464}, 1021-1024 (2010).

\bibitem{Ek}A. K. Ekert, Quantum cryptography based on Bell's theorem, \textit{Phys. Rev. Lett.} {\bf 67}, 661 (1991).

\bibitem{SG} V. Scarani and N. Gisin, Quantum communication between $N$ partners and Bell's inequalities, \textit{Phys. Rev. Lett}. \textbf{87}, 117901 (2001).

\bibitem{MY}D.  Mayers and A.  Yao, Quantum cryptography with imperfect apparatus. {\it Proc. of the 39th IEEE Symposium on Foundations of Computer Science} (IEEE Computer Society, Los Alamitos, 1998), p. 503.

\bibitem{AGM}A. Ac\'{\i}n,  N. Gisin, and L. Masanes, From Bell's theorem to secure quantum key distribution, \textit{Phys. Rev. Lett.} \textbf{97}, 120405 (2006).

\bibitem{ABG}A. Ac\'{\i}n, N. Brunner, N. Gisin, S. Massar, S. Pironio, and V. Scarani, Device-independent security of quantum cryptography against collective attacks, \textit{Phys. Rev. Lett.} \textbf{98}, 230501 (2007).


\bibitem{BPA}N. Brunner, S. Pironio, A. Ac\'{\i}n, N. Gisin, A. A. Methot, and V. Scarani, Testing the dimension of Hilbert spaces, \textit{Phys. Rev. Lett.} \textbf{100}, 210503 (2008).

\bibitem{BCMD}H.  Buhrman, R. Cleve, S. Massar, and R. de Wolf, Nonlocality and communication complexity, \textit{Rev. Mod. Phys.} {\bf 82}, 665 (2010).

\bibitem{BCPS}N. Brunner, D. Cavalcanti, S. Pironio, V. Scarani, and S. Wehner, Bell nonlocality, \textit{Rev. Mod. Phys.} \textbf{86}, 419-478 (2014).


\bibitem{HHH0}R. Horodecki, P. Horodecki, M. Horodecki, and K. Horodecki, Quantum entanglement, \textit{Rev. Mod. Phys}. \textbf{81}, 865 (2009)

\bibitem{CSS}D. Cavalcanti, P. Skrzypczyk, and I. \v{S}upi\'{c}, All entangled states can demonstrate nonclassical teleportation, \textit{Phys. Rev. Lett.} \textbf{119}, 110501 (2017).

\bibitem{ZZHE}M. Zukowski, A. Zeilinger, M. A. Horne, and A. K. Ekert, ``Event-ready-detectors" Bell experiment via entanglement swapping,  \textit{Phys. Rev. Lett.} \textbf{71}, 4287 (1993).

\bibitem{SPG} A. J. Short, S. Popescu, and N. Gisin, Entanglement swapping for generalized nonlocal correlations, \textit{Phys. Rev. A} \textbf{73}, 2518-2521 (2005).

\bibitem{TRO}P. R. Tapster, J. G. Rarity, and P. C. M. Owens, Violation of Bell's inequality over 4 km of optical fiber, \textit{Phys. Rev. Lett}. \textbf{73}, 1923 (1994).

\bibitem{TBZ}W. Tittel, J. Brendel, H. Zbinden, and N. Gisin, Violation of Bell inequalities by photons more than 10 km apart, \textit{Phys. Rev. Lett.} \textbf{81}, 3563 (1998).

\bibitem{AB}M. Aspelmeyer, H. R. B\"{o}hm, T. Gyatso, T. Jennewein, R. Kaltenbaek, M. Lindenthal, G. Molina-Terriza, A. Poppe, K. Resch, M. Taraba, R. Ursin, P. Walther, and A. Zeilinger, Long-distance free-space distribution of quantum entanglement, \textit{Science} \textbf{301}, 621-623 (2003).

\bibitem{YC}J. Yin, Y. Cao, Y.-H. Li, S.-L. Liao, L. Zhang, J.-G. Ren, W.-Q. Cai, W.-Y. Liu, B. Li, H. Dai, G.-B. Li, Q.-M. Lu, Y.-H. Gong, Y. Xu, S.-L. li, F.-Z. Li, Y.-Y. Yin, Z.-Q. Jiang, M. Li, J.-J. Jia, G. Ren, D. He, Y.-L. Zhou, X.-X. zhang, N. Wang, X. Chang, Z.-C. Zhu, N.-L. Liu, Y.-A. Chen, C.-Y. Lu, R. Shu, C.-Z. Peng, J.-Y. Wang, and J.-W. Pan, Satellite-based entanglement distribution over 1200 kilometers, \textit{Science} \textbf{356}, 1140-1144 (2017).

\bibitem{Kimb}H. J. Kimble, The quantum Internet, \textit{Nature} (London) \textbf{453}, 1023 (2008).

\bibitem{Ritt}S. Ritter, C. N\"{o}lleke, C. Hahn, A. Reiserer, A. Neuzner, M. Uphoff, M. M\"{u}cke, E. Figueroa, J. Bochmann, and G. Rempe, An elementary quantum network of single atoms in optical cavities, \textit{Nature} \textbf{484}, 195-200 (2012).

\bibitem{BDCZ}H.-J. Briegel, W. D\"{u}r, J. I. Cirac, and P. Zoller, Quantum repeaters: The role of imperfect local operations in quantum communication, \textit{Phys. Rev. Lett.} \textbf{81}, 5932 (1998).

\bibitem{DLCZ}L.-M. Duan, M. Lukin, J. I. Cirac, and P. Zoller,  Long-distance quantum communication with atomic ensembles and linear optics, \textit{Nature} \textbf{414}, 413 (2001).

\bibitem{ATL}K. Azuma, K. Tamaki, and H.-K. Lo, All-photonic quantum repeaters, \textit{Nat. Comm.} \textbf{6}, 6787 (2015).

\bibitem{ZPD}M. Zwerger, A. Pirker, V. Dunjko, H. J. Briegel, and W. D\"{u}r, Long-range big quantum-data transmission, \textit{Phys. Rev. Lett.} \textbf{120}, 030503 (2018).

\bibitem{Cire} B. S. Cirel'son, Quantum generalizations of Bell's inequality, \textit{Lett. Math. Phys.} \textbf{4}, 93 (1980).

\bibitem{ABH}J.-M. A. Allen, J. Barrett, D. C. Horsman, C. M. Lee, and R. W. Spekkens, Quantum common causes and quantum causal models, \textit{Phys. Rev. X} \textbf{7}, 031021 (2017).

\bibitem{CAS}D. Cavalcanti, M. L. Almeida, V. Scarani, and A. Ac\'{\i}n, Quantum networks reveal quantum nonlocality, \textit{Nat. Comm}. \textbf{2}, 184 (2011).

\bibitem{BGP}C. Branciard, N. Gisin, and S. Pironio, Characterizing the nonlocal correlations created via entanglement swapping, \textit{Phys. Rev. Lett.} \textbf{104}, 170401 (2010).

\bibitem{GMT}N. Gisin, Q. Mei, A. Tavakoli, M. O. Renou, and N. Brunner, All entangled pure quantum states violate the bilocality inequality, \textit{Phys. Rev. A} \textbf{96}, 020304(R) (2017).

\bibitem{ACS}F. Andreoli, G. Carvacho, L. Santodonato, R. Chaves, and F. Sciarrino, Maximal qubit violation of $n$-locality inequalities in a star-shaped quantum network,  \textit{New J. Phys.} \textbf{19}, 113020 (2017).

\bibitem{LS}R. Chaves, L. Luft, and D. Gross, Causal structures from entropic information: Geometry and novel scenarios, \textit{New J. Phys.} \textbf{16}, 043001 (2014).

\bibitem{CH} R. Chaves, Polynomial Bell inequalities, \textit{Phys. Rev. Lett.} \textbf{116}, 010402 (2016).

\bibitem{RBB}D. Rosset, C. Branciard, T. J. Barnea, G. P\"{u}tz, N. Brunner, and N. Gisin, Nonlinear Bell inequalities tailored for quantum networks, \textit{Phys. Rev. Lett.} \textbf{116}, 010403 (2016).

\bibitem{Luo}M.-X. Luo, Computationally efficient nonlinear Bell inequalities for quantum networks, \textit{Phys. Rev. Lett.} \textbf{120}, 140402 (2018).

\bibitem{GHZ} D. M. Greenberger, M. A. Horne, and A. Zeilinger, in Bell's Theorem, Quantum Theory and Conceptions of the Universe, edited by M. Kafatos (Kluwer, Dordrecht, 1989), p. 69.

\bibitem{CL}A. Cabello and  J.-A. Larsson, Minimum detection efficiency for a loophole-free atom-photon Bell experiment, \textit{Phys. Rev. Lett}. \textbf{98}, 220402 (2007).

\bibitem{BG}N. Brunner, N. Gisin, V. Scarani, and C. Simon, Detection loophole in asymmetric Bell experiments, \textit{Phys. Rev. Lett.} \textbf{98}, 220403 (2007).

\bibitem{TWE}T. R. Tan, Y. Wan, S. Erickson, P. Bierhorst, D. Kienzler, S. Glancy, E. Knill, D. Leibfried, and D. J. Wineland, Chained Bell inequality experiment with high-efficiency measurements, \textit{Phys. Rev. Lett.} \textbf{118}, 130403 (2017).

\bibitem{SBB}D. J. Saunders, A. J. Bennet, C. Branciard, and G. J. Pryde, Experimental demonstration of nonbilocal quantum correlations, \textit{Sci. Adv.} \textbf{3}, 1602743 (2017).

\bibitem{CASB} G. Carvacho, F. Andreoli, L. Santodonato, M. Bentivegna, R. Chaves, and F. Sciarrino, Experimental violation of local causality in a quantum network, \textit{Nat. Comm.} \textbf{8}, 14775 (2017).

\bibitem{Luo1}M.-X. Luo, Typical Werner states satisfying all linear Bell inequalities with dichotomic measurements, \textit{Phys. Rev. A} \textbf{97}, 042301 (2018).

\bibitem{HHH}M. Horodecki, P. Horodecki, and R. Horodecki, Inseparable two spin-(1/2) density matrices can be distilled to a singlet form, \textit{Phys. Rev. Lett.} \textbf{78}, 574-577 (1997).

\bibitem{SI}See Supplemental Material which includes Refs.[3-5,7,11,12,38,42,47,55], for the detailed proofs of Theorems 1, 2, and 3.

\bibitem{BLW}N. Biggs, E. Lloyd, and R. Wilson, \textit{Graph Theory}, Oxford University Press, 1986.

\bibitem{SS}M. Seevinck \& G. Svetlichny, Bell-type inequalities for partial separability in $N$-particle systems and quantum mechanical violations, \textit{Phys. Rev. Lett.} \textbf{89}, 060401 (2002).

\bibitem{CGP} D. Collins,  D. Gisin,  S. Popescu,  D. Roberts, \&  V. Scarani, Bell-type inequalities to detect true $n$-Body nonseparability, \textit{Phys. Rev. Lett.} \textbf{88}, 170405 (2002).

\bibitem{Sve}D. Svetlichny, Distinguishing three-body from two-body nonseparability by a Bell-type inequality, \textit{Phys. Rev. D} \textbf{35}, 3066 (1987).

\bibitem{CHW}E. G. Cavalcanti, M. J. W. Hall, and H. M. Wiseman, Entanglement verification and steering when Alice and Bob cannot be trusted, \textit{Phys. Rev. A} \textbf{87}, 032306 (2013).

\bibitem{BRL} C. Branciard,  D. Rosset,  Y. C. Liang, \&  N. Gisin,  Measurement-device-independent entanglement witnesses for all entangled quantum states, \textit{Phys. Rev. Lett.} \textbf{110}, 060405 (2013).

\bibitem{Deng}D.-L. Deng, Machine learning Bell nonlocality in quantum many-body systems, arXiv:1710.04226 (2017).

\bibitem{LGLS}Y. Li, M. Gessner, W. Li, and A. Smerzi, Hyper- and hybrid nonlocality, \textit{Phys. Rev. Lett.} \textbf{120}, 050404 (2018).

\bibitem{ACTC}B. Amaral, A. Cabello, M. T. Cunha, and L. Aolita, Noncontextual wirings, \textit{Phys. Rev. Lett.} \textbf{120}, 130403 (2018).

\end{thebibliography}

\begin{thebibliography}{99}
\bibitem{SQ}F. Buscemi, All entangled quantum states are nonlocal, \textit{Phys. Rev. Lett.} \textbf{108}, 200401 (2012).

\bibitem{Bell}J. S. Bell, On the Einstein-Podolsky-Rosen paradox, \textit{Phys}. \textbf{1}, 195 (1964).

\bibitem{PR}S. Popescu and  D. Rohrlich, Generic quantum nonlocality, \textit{Phys. Lett. A} \textbf{166}, 293 (1992).

\bibitem{Cire} B. S. Cirel'son, Quantum generalizations of Bell's inequality, \textit{Lett. Math. Phys.} \textbf{4}, 93 (1980).

\bibitem{HHH}M. Horodecki, P. Horodecki, and R. Horodecki, Inseparable two spin-(1/2) density matrices can be distilled to a singlet form, \textit{Phys. Rev. Lett.} \textbf{78}, 574-577 (1997).

\bibitem{GMT}N. Gisin, Q. Mei, A. Tavakoli, M. O. Renou, and N. Brunner, All entangled pure quantum states violate the bilocality inequality, \textit{Phys. Rev. A} \textbf{96}, 020304(R) (2017).

\bibitem{Luo}M.-X. Luo, Computationally efficient nonlinear Bell inequalities for quantum networks, \textit{Phys. Rev. Lett.} \textbf{120}, 140402 (2018).

\bibitem{CHSH}J. F. Clauser, M. A. Horne, A. Shimony, and R. A. Holt, Proposed experiment to test local hidden-variable theories, \textit{Phys. Rev. Lett.} \textbf{23}, 880 (1969).

\bibitem{Gisin1}N. Gisin, Bell's inequality holds for all non-product states, \textit{Phys. Lett. A} \textbf{154}, 201 (1991).

\bibitem{YCZ}S. Yu, Q. Chen, C. Zhang, C. H. Lai, and C. H. Oh, All entangled pure states violate a single Bell's inequality, \textit{Phys. Rev. Lett}. \textbf{109}, 120402 (2012).

\end{thebibliography}
\end{document}